\documentclass[fleqn,usenatbib]{mnras}
\usepackage{lmodern}
\usepackage{newtxtext,newtxmath}

\usepackage[T1]{fontenc}
\usepackage{fix-cm}

\DeclareRobustCommand{\VAN}[3]{#2}
\let\VANthebibliography\thebibliography
\def\thebibliography{\DeclareRobustCommand{\VAN}[3]{##3}\VANthebibliography}

\usepackage{subcaption}

\usepackage{graphicx}	
\usepackage{amsmath}	
\usepackage[utf8]{inputenc}
\usepackage{newunicodechar}
\newunicodechar{−}{-}

\title[Quantifying bar deceleration with resonance]{Twisted resonant orbits: Quantifying the deceleration of the Galactic bar}

\author[A. Archer et al.]{
Alice Archer,$^{1}$\thanks{E-mail: aliceeaarcher@icloud.com}
Adam Dillamore,$^{2,3}$
Hanyuan Zhang$^{1}$
and Vasily Belokurov$^{1}$
\\
$^{1}$Institute of Astronomy, University of Cambridge, Madingley Road, Cambridge CB3 0HA, UK\\
$^{2}$School of Mathematics and Physics, University of Surrey, Guildford GU2 7XH, UK\\
$^{3}$Department of Physics and Astronomy, University College London, London, WC1E 6BT, UK
}

\date{Accepted XXX. Received YYY; in original form ZZZ}

\pubyear{\the\year{}}

\begin{document}
\label{firstpage}
\pagerange{\pageref{firstpage}--\pageref{lastpage}}
\maketitle

\begin{abstract}
Recent structural discoveries in Gaia Data Release 3 reveal distinct, chevron-shaped over-densities in the radial phase space of the halo. Although initial interpretations identify these features as the phase-mixed relics of a past galaxy merger, their prominent asymmetry and persistence at high metallicities strongly challenge this interpretation. These features are better explained as a dynamical response to a Galactic bar, where stars are trapped in resonances. We establish the fractional $v_r$ asymmetry of the chevrons as a quantitative diagnostic tool for the bar's rotational state. We utilise test-particle simulations of a Milky Way-like galaxy to model various bar pattern speed evolutions, including decelerating and constant speed scenarios. By mapping the fractional asymmetry of the phase-space chevrons across a parameter space of bar deceleration rates and pattern speeds, we establish a direct correlation between the bar's deceleration and the rotational response of the resonant orbits. Comparing these simulated constraints with observational data from Gaia DR3 provides a framework for measuring the current deceleration of the Galactic bar, and a probe of the amount of dynamical friction occurring between the baryonic and dark matter content of the Milky Way. We find a dimensionless deceleration parameter $\eta \equiv - \dot{\Omega}_b/\Omega_b^2 = 0.0015^{+0.0016}_{-0.0009}$, equivalent to $\dot{\Omega}_b = -2.23^{+1.50}_{-2.46} \, \text{km s}^{-1} \text{kpc}^{-1} \text{Gyr}^{-1}$, in agreement with previous estimates obtained with independent methods. Gaia DR4 should sharpen this measurement considerably.

\end{abstract}




\section{Introduction}

The publication of Gaia Data Release 3 (DR3) \citep{2016A&A...595A...1G, 2021A&A...649A...1G} provides a dramatic increase in the number of stars with full 6D phase-space information, significantly advancing our understanding of the Milky Way. One discovery from this dataset was a set of clear over-densities of stars in chevron-shaped regions of $r-v_r$ phase-space at low angular momentum. 

These features were identified in the data and initially hypothesised to be the phase-mixed relics of a past galaxy merger \citep{2023MNRAS.518.6200B}, so-called `energy wrinkles'. However, the prominence, asymmetry and persistence of these structures even in stars with relatively high metallicity ($[\text{Fe}/\text{H}] > -  0.7$) \citep{2023MNRAS.518.6200B} challenged this hypothesis. An alternative explanation, which fits the morphology of the data, posits that these features are instead a dynamical response to the Galactic bar. 

The presence of the bar is well-established, with a consensus on its existence forming by the mid-1990s \citep{1991MNRAS.252..210B, 1994ApJ...425L..81W}. While resonant perturbation theory has long been used to explain substructure in the Galactic disc, such as the Hyades and Sirius streams \citep{1998AJ....115.2384D}, its effect on the stellar halo is a more recent focus \citep{Dillamore_2023, Dillamore2025}. 

The Galactic bar may also be decelerating. This is a consequence of dynamical friction exerted by the Milky Way's dark matter halo \citep{1984MNRAS.209..729T, Debattista_2000, Athanassoula2003, Weinberg2007}. The deceleration rate of the Galactic bar is a probe of the dark halo's presence in the inner Galaxy; therefore, knowing this deceleration rate is foundational to understanding the Milky Way, as well as a useful value for future Galactic dynamics studies.

\begin{figure*}
    \centering
    \includegraphics[width=\linewidth]{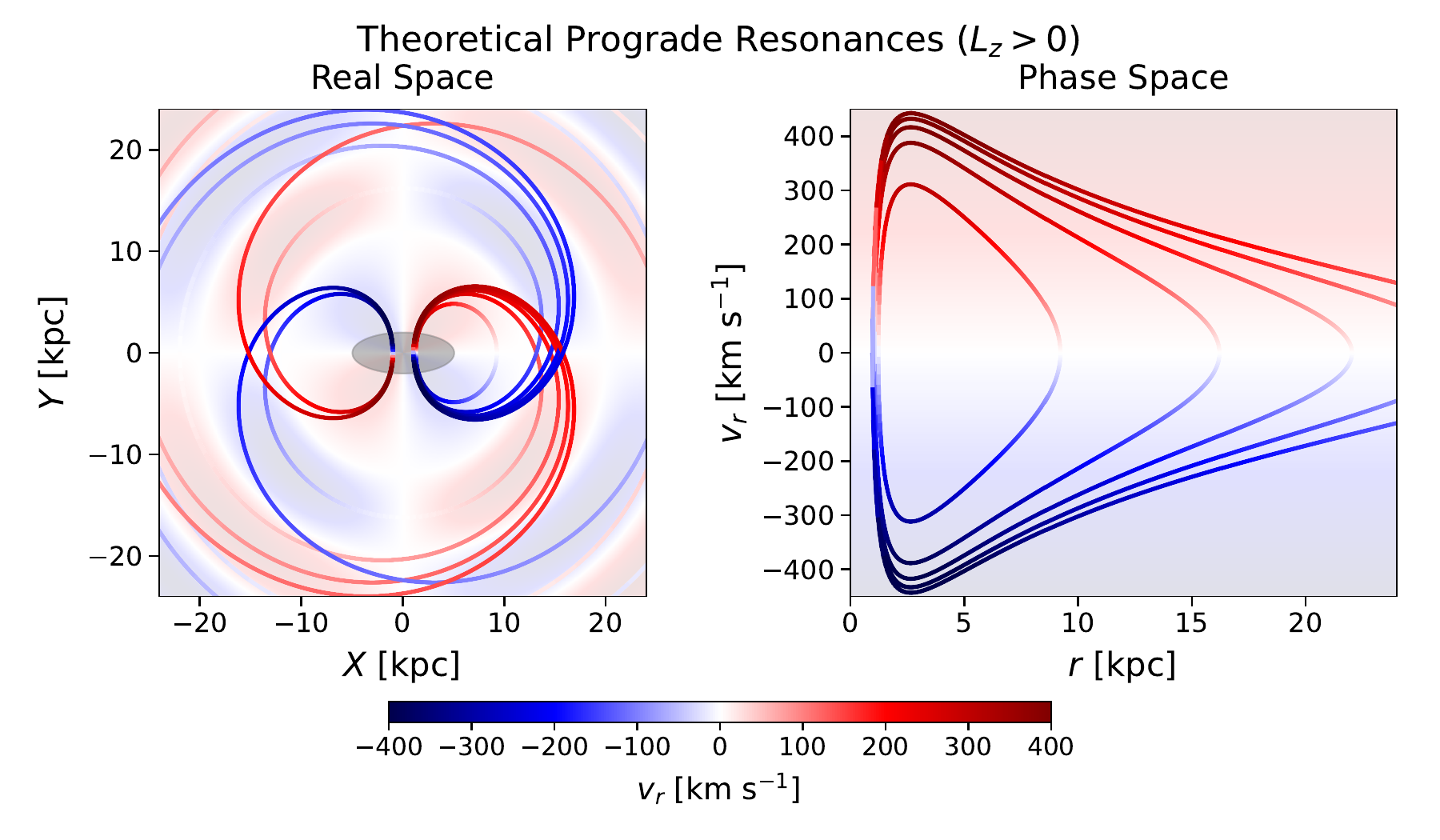}
        \caption{Theoretical distribution of prograde resonant orbits ($L_z>0$) in the Galactic plane and corresponding radial phase-space, coloured by Galactocentric radial velocity $v_r$. The left panel shows the spatial distribution of the resonant orbits in Cartesian coordinates, while the right panel shows their corresponding distribution in $(r,v_r)$ phase-space. The coherent radial motions of the trapped stars produce distinct chevron-shaped structures, illustrating how resonant orbital dynamics generate the observed phase-space over-densities. Based on the theoretical framework presented in \citet{Dillamore_2023}.}
    \label{fig:theoretical_orbits} 
\end{figure*}

It was hypothesised that chevrons are the manifestation of halo stars trapped in resonant orbits \citep{2024MNRAS.532.4389D}. Under this model, the bar effectively sweeps up stars as it rotates. The theoretical framework for this mechanism was established as a consequence of a decelerating bar \citep{Chiba2021}. This framework is consistent with modern estimates favouring a slower and longer bar \citep{Sormani_2015}. Resonance occurs when a linear combination (with rational coefficients) of a star's intrinsic orbital frequencies matches the rotational pattern speed of the bar, $\Omega_b$ \citep{Binney}. Trapped halo stars form symmetric orbital paths in the rotating frame of the bar. By isolating the radial component of this motion, distinct regions of the Galaxy are revealed to be dominated by either outward ($v_r > 0$) or inward ($v_r < 0$) bulk flow. In $r-v_r$ phase-space, this coherent radial motion carves out distinct chevron shapes. A visualisation of these structures is shown in Figure \ref{fig:theoretical_orbits} which is based on the work laid out in \citet{Dillamore_2023}. These bar-formed chevrons appear in simulations of a Milky Way-like galaxy \citep{2024MNRAS.532.4389D} and align with the observational data more closely than those produced by mergers. 

Crucially, in the rotating frame, stars experience an additional Euler force proportional to $\dot{\Omega}_b$. To remain trapped, stars along resonant orbits are forced to shift their orbital phase. This offsets the $v_r$-asymmetry out of alignment with the bar's major axis, twisting the chevron signature \citep{Chiba2021}.

The primary goal of this work is to establish this spatial asymmetry as a quantitative diagnostic tool for the motion of the Galactic bar. If phase-space chevrons are indeed a dynamical response to the bar, their morphology directly encodes its rotational history. Building on previous work that links decelerating bars to resonant orbits \citep{Chiba2021, 2024MNRAS.532.4389D,Dillamore_2023}, we demonstrate that the shift caused by the Euler force provides a measurable signature of the system's deceleration. 

To quantify this effect, we utilise test-particle simulations to model the chevron asymmetry across a variety of Milky Way-like configurations. By mapping the rotational deviation of the kinematic signature, we create a diagnostic framework for the deceleration the bar is currently experiencing. We then compare these simulated constraints directly against observational data from Gaia DR3 to test modern literature estimates of the dimensionless deceleration parameter $\eta \equiv - \frac {\dot{\Omega}_b}{\Omega_b^2}= 0.0031 \pm 0.0008$ \citep{Chiba2021}, an estimate verified by \citet{Zhang_2025} using the age-metallicity sequences of the disc.

Section \ref{math_framework} introduces the mathematical framework required for the rest of the analysis. Section \ref{sims} describes the analysis and results of our simulated datasets. In particular, Section \ref{N-body} describes the creation of our simulated data, Section \ref{sims results} the subsequent analysis, and Section \ref{mock} details the evolution from the raw data into a mock catalogue. Section \ref{mcmc_fitting_section} describes the measurement of deviation across both datasets. Section \ref{methods} shows how we reproduce and extend the analysis on the Gaia data, outlining our methodological framework. In particular, Section \ref{gaia methods} details the processing of the Gaia data, and then Sections \ref{gaiachevrons} to \ref{fin_dev} detail the analysis of this data, which is then directly compared to the simulated data in Section \ref{observation results}. The results are summarised and discussed in Section \ref{discussion}. Finally, our conclusions are made in Section \ref{conclusion}.

\section{Mathematical Framework} \label{math_framework}

In this study, we define a reference frame where the Galactic bar rotates anti-clockwise (i.e. with positive angular momentum, $L_z>0$). We transform the observational Gaia data to match this reference frame as described in Section \ref{transformation}. 

The bar acts as a rotating, non-axisymmetric perturbation to the gravitational potential \citep{Binney}. Its physical morphology is tied to its rotational kinematics: a rapidly rotating system maintains a physically shorter bar, confining its associated resonant structures to smaller galactic radii \citep{Sormani_2015}. Consequently, a decelerating bar naturally extends these chevron signatures outward over time.

To mathematically define the resonant orbits that generate these chevrons, we restrict our analysis to planar orbits. The resonance condition (equation \ref{resonant_eq} below) involves only the radial and azimuthal frequencies. Our simulations integrate full three-dimensional orbits, and the chevrons and their asymmetry are clearly retained after projection onto the Galactic plane, so we analyse $(R, \phi, v_R)$ only.
 
 
The exact resonance condition is given by:
\begin{equation} \label{resonant_eq}
    l\Omega_R + m(\Omega_\phi - \Omega_b) = 0 \quad l,m \in \mathbb{Z}
\end{equation}
where $\Omega_R$ is the frequency of radial oscillations and $\Omega_{\phi}$ is the frequency of azimuthal revolutions of the given star \citep{2024MNRAS.532.4389D}. Because opposite points in the Galaxy ($\phi \rightarrow \phi + \pi$) present an identical gravitational environment relative to the spinning bar, the trapped stars form a pattern which is periodic in azimuth. Practically, this means opposite points in the galaxy behave the same and can be superimposed on top of each other to increase the number of data points. 

The shift due to the Euler force was derived by \citet{Chiba2021}. We quantify the deceleration using the dimensionless parameter: 
\begin{equation} \label{eta}
\eta \equiv -\frac{\dot\Omega_b}{\Omega_b^2}
\end{equation}
The kinematic behaviour of the trapped stars depends strictly on the ratio $\eta/A_b$, where $A_b$ is proportional to the bar's strength. For a constant-speed bar ($\eta=0$), the equilibrium centres of the resonant orbits sit at $\sin \theta_{\mathrm{eq}} = 0$, meaning the kinematic signature perfectly aligns with the bar's major axis \citep{Chiba2021}. Such an arrangement is illustrated in Figure \ref{fig:theoretical_orbits}.

However, for a decelerating bar ($\dot{\Omega}_b < 0 \implies \eta > 0$), because the circular frequency $\Omega_{\phi}$ generally decreases with galactic radius, the resonance sweeps outwards as time passes. To remain in balance against the Euler force, trapped orbits are forced to a new equilibrium, approximately given by \citep{Chiba2021}:
\begin{equation} \label{shifted eq phase}
    \sin \theta_{\mathrm{eq}} = \frac{\eta}{A_b} 
\end{equation}
As long as the deceleration remains slow relative to the trapping strength ($|\eta/A_b| < 1$), the resonance island remains intact. The stars undergo adiabatic `resonant dragging' to larger radii, and the macroscopic chevron signature is offset azimuthally. If the deceleration is too rapid ($|\eta/A_b| > 1$), this adiabatic trapping region vanishes, freeing the stars and destroying the chevron signature entirely \citep{Chiba2021}.

We define the deviation $\Delta \phi_{\odot}$ as the azimuth, measured from the bar major axis in the direction of rotation, at which the $m=2$ component of the radial-velocity asymmetry in the Solar annulus ($R_\odot \pm 1.25$ kpc) changes sign from negative to positive (Section \ref{mcmc_fitting_section}). For a bar rotating at constant speed, and in the absence of observational errors, $\Delta \phi_{\odot} = 0$; the Euler force in a decelerating bar shifts it to $\Delta \phi_{\odot} > 0$.

\section{Simulated Data} \label{sims}

\subsection{Test Particle Simulations} \label{N-body}

We use test-particle simulations of a Milky Way-like galaxy. The constant-speed runs last 8 Gyr. All decelerating runs start at $\Omega_b = 60 \, \text{km s}^{-1} \text{kpc}^{-1}$ and end at $\Omega_b \approx 25 \, \text{km s}^{-1} \text{kpc}^{-1}$, so their duration depends on $\eta$, from $\approx 7$ Gyr for $\eta = 0.004$ to $\approx 260$ Gyr for $\eta = 0.0001$. The simulation generates $N = 2 \times 10^7$ particles sampled from an action-based Double Power Law distribution function \citep{Posti_2015}, implemented via the \textsc{Agama} framework \citep{Vasiliev_2018}. Specifically, this halo distribution is defined by an inner slope of $\Gamma = 1$, an outer slope of $\beta = 4$, and a break action of $J_0 = 2000$. The simulation then integrates their orbits independently in a pre-determined time-dependent potential. 

This potential is a composite of a static axisymmetric background (representing the Milky Way's disc and dark matter halo) and a superimposed rotating, non-axisymmetric bar. Both components are based on the \citet{Hunter_2024} model. 

The bar potential is introduced smoothly, with its amplitude growing between times $t_0$ and $t_1$ via a quintic polynomial step function. Following this initial growth phase, the pattern speed $\Omega_b(t)$ evolves according to a continuous function parametrised by the deceleration rate $\eta$. Crucially, to ensure a controlled dynamical comparison, the intrinsic strength of the bar is held constant across all simulations. However, the physical morphology of the bar is dynamic; it elongates as the pattern speed decreases. This ensures that the bar length varies consistently throughout a simulation's lifetime.

This model provides a realistic description of the Milky Way's bulge, disc, and dark matter halo distributions. These simulations are adapted from those run in \citet{2024MNRAS.532.4389D}.

We run four types of galactic simulations:
\begin{enumerate}
    \item a decelerating bar,
    \item a bar with constant rotation speed,
    \item a bar which decelerates and then settles at a constant speed (the results of which are discussed in Appendix \ref{decelstop}),
    \item and an accelerating bar (the results of which are discussed in Appendix \ref{accel}).
\end{enumerate}


    
   
    

We expect the chevrons' appearance to hold information about the current speed and deceleration. By running the decelerating simulations at a set of different deceleration rates ($\eta$) we can get a comprehensive picture of the chevrons across a physically motivated region of deceleration-speed space, where $\eta$ is defined in equation \ref{eta}.

\subsubsection{Decelerating Bar}
 
In the case of a decelerating bar, the evolution of the pattern speed $\Omega_b(t)$ in the simulation follows different regimes depending on the time $t$:

\begin{equation}
\Omega_b(t) = 
\begin{cases} 
\Omega_{b,1} & , t \leq t_1 \\
\Omega_{b,1} \left[ 1 + \frac{1}{2} \eta \Omega_{b,1} \frac{(t - t_1)^2}{t_2 - t_1} \right]^{-1} & , t_1 < t \leq t_2 \\
\Omega_{b,2} [ 1 + \eta \Omega_{b,2} (t - t_2) ]^{-1} & , t > t_2
\end{cases}
\end{equation}

where
\begin{equation*}
\Omega_{b,2} = \Omega_{b,1} \left[ 1 + \frac{1}{2} \eta \Omega_{b,1} (t_2 - t_1) \right]^{-1} .
\end{equation*} 

This profile allows the pattern speed to decrease from an initial $\Omega_b = 60 \, \text{km s}^{-1} \text{kpc}^{-1}$ down to $\Omega_b \approx 25 \, \text{km s}^{-1} \text{kpc}^{-1}$, to encompass a generous range of physically motivated values.

Initially, the deceleration parameter $\eta$ is varied from $0.001$ up to $0.016$ in steps of $0.001$. This is chosen to comfortably bracket the modern literature estimate of $\eta = 0.0031 \pm 0.0008$ \citep{Chiba2021}. However, at higher deceleration rates ($\eta \geq 0.004$), the kinematic signature effectively disappears due to the streaming of stars out of the resonant orbits. Since the signal decays at these higher values and the central values of literature estimates of the Milky Way's deceleration rate lie below this threshold \citep{Chiba2021, Zhang_2025}, we restrict our primary analysis to $0<\eta\leq0.004$. We also use finer increments of $0.0001$ to allow for a more precise analysis of how $\Delta \phi_{\odot}$ (the deviation defined in Section \ref{math_framework}) varies with deceleration. 

To allow for direct comparisons between snapshots, we store $26$ snapshots at regular intervals between $t_1$ and $t_f$. The speed profiles all have the same shape (except for being compressed in time) so taking these intervals results in each $\eta$ having a set of snapshots with the same set of pattern speeds. The speeds range from $60$ to $24.9 \text{ km s}^{-1} \text{kpc}^{-1}$. This leaves us with a set of snapshots with varying pattern speed and deceleration rate to analyse. Runs with small $\eta$ therefore span long times, but because the phase response to a change in the deceleration damps on a time-scale of $\sim 3$ Gyr (Appendix \ref{decelstop}), $\Delta \phi_{\odot}$ in any snapshot is set mainly by the last few Gyr of evolution.

\subsubsection{Bars with Constant Speed}

A set of simulations with $\eta=0$ is also generated. For each simulation, we hold the pattern speed constant for $8 \, \text{Gyr}$, where the first billion years are the bar growth phase. Each simulation is run at a different pattern speed. The set of speeds is the same as those sampled for the decelerating models, providing a direct comparison.

\subsection{Simulated Data Processing} \label{sims results}

\subsubsection{Phase-Space Chevrons} \label{phase space chevrons}

We generate the ($r$-$v_r$) phase-space data by transforming the 6D phase-space data from Cartesian coordinates centred at the galactic centre (GC) into cylindrical polar coordinates centred at the GC. In each snapshot the data is rotated such that the major bar axis lies across the $\phi=0^{\circ}$, $\phi=180^{\circ}$ line. The position of the Sun is taken to be $r=8.122 \text{ kpc}$ \citep{GRAVITY:2018ofz} and $\phi=333^{\circ}$ \citep{Wegg_2015}, where the system rotates anti-clockwise. To isolate the in-plane kinematics, we neglect the vertical $z$ dimension.

For each azimuthal slice the snapshot data is separated into two groups: stars with positive and negative angular momentum. There is a cut limiting the magnitude of $L_z$ to below $500$ kpc km s$^{−1}$, this excludes stars on disc-like orbits and bounds our analysis to the relevant phase-space regions.

Next, to analyse the azimuthal dependence of the resonant kinematics, we partition both populations into four spatial quadrants each. Because the opposite sides of the Galaxy are in equivalent potentials with respect to the bar (so behave identically) we superimpose the data from opposing quadrants to maximise the signal-to-noise ratio of our phase-space maps. 

\begin{figure}
    \centering
    \begin{subfigure}[b]{\columnwidth}
        \centering
        \includegraphics[width=\linewidth]{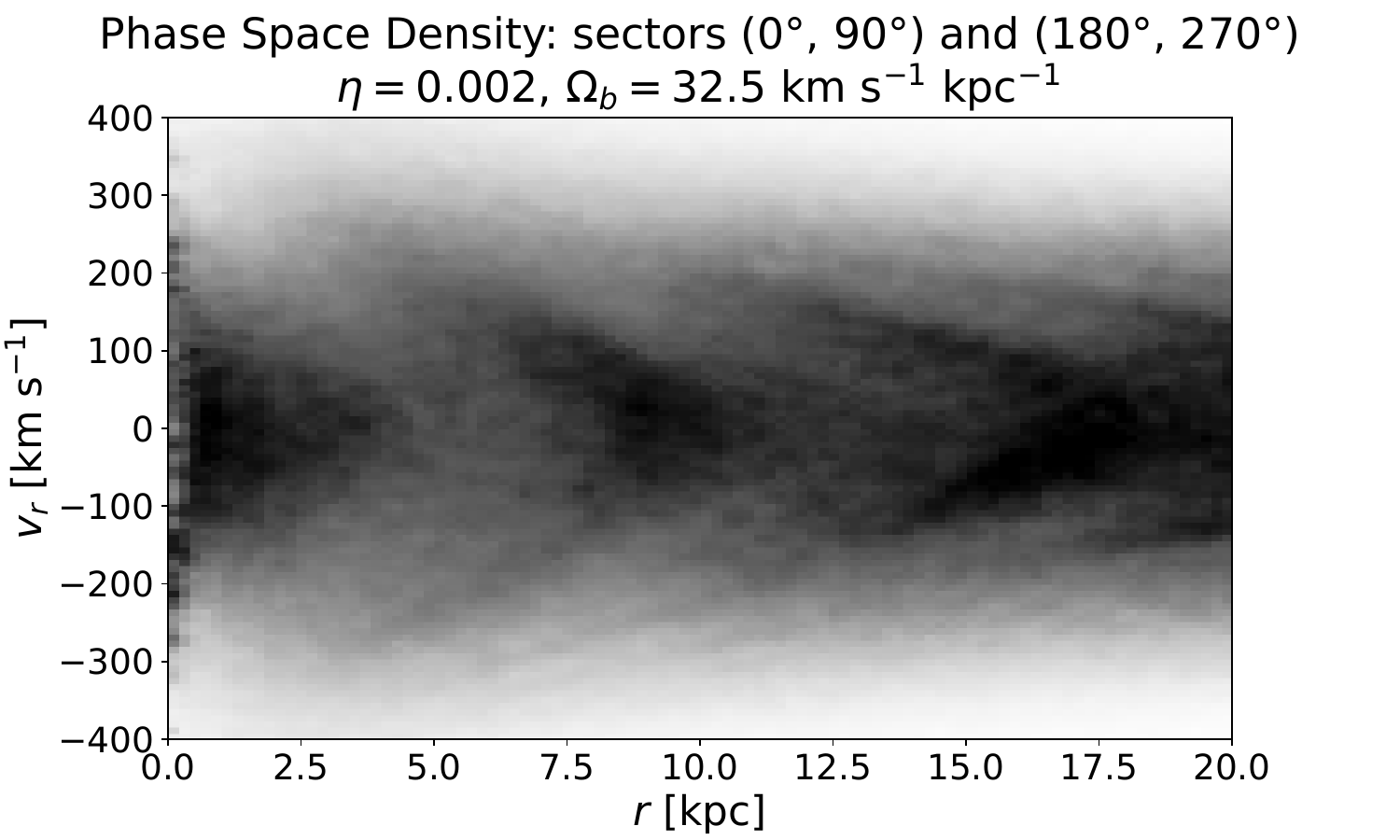}
        \caption{$(0^{\circ}, 90^{\circ})\; \& \; (180^{\circ}, 270^{\circ})$}
        \label{fig:chevrons_pos_45}
    \end{subfigure}
    \hfill
    \begin{subfigure}[b]{\columnwidth}
        \centering
        \includegraphics[width=\linewidth]{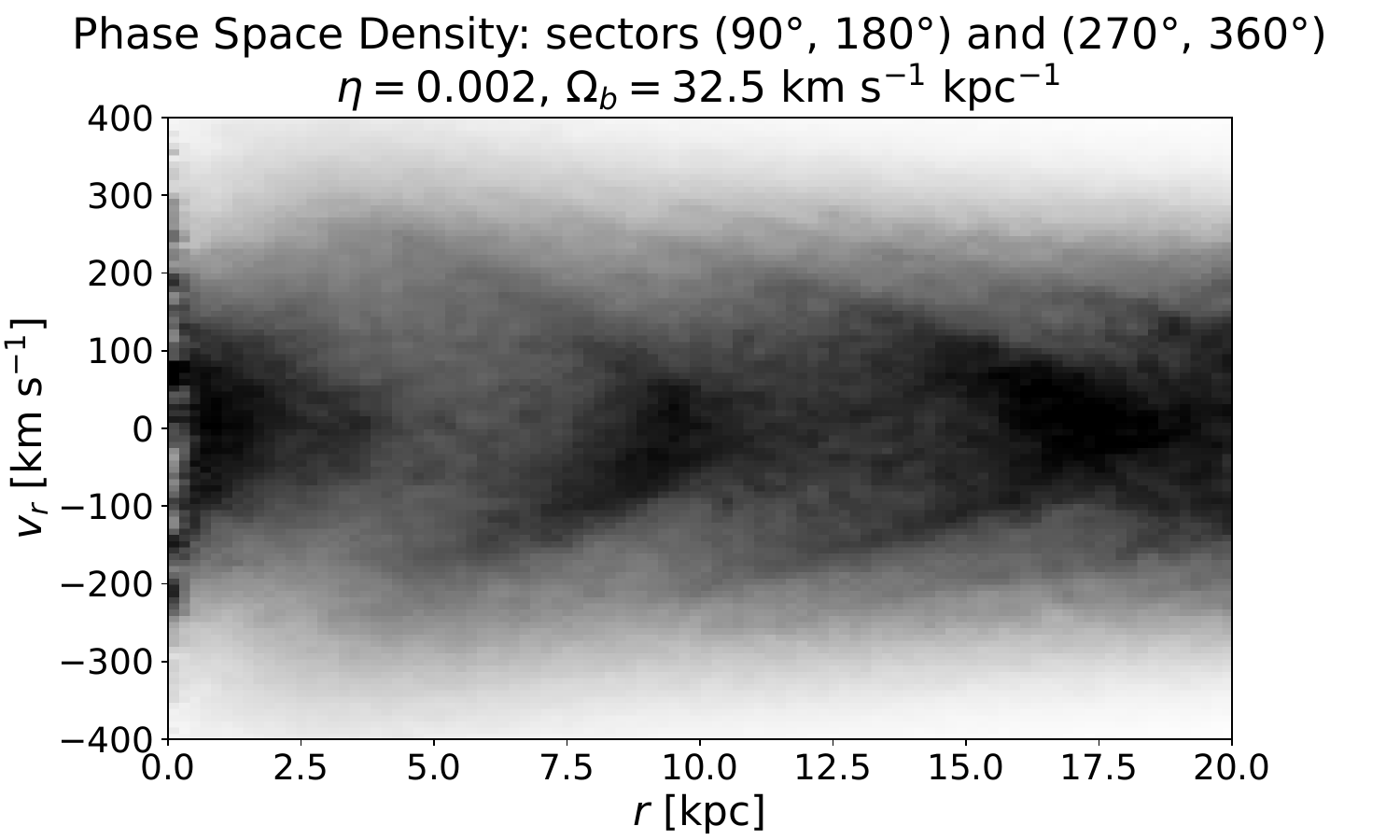}
        \caption{$(90^{\circ}, 180^{\circ})\; \& \; (270^{\circ}, 360^{\circ})$}
        \label{fig:chevrons_pos_135}
    \end{subfigure}

    \caption{Simulated radial phase-space distributions for prograde particles with $0<L_z<500\,\mathrm{kpc\,km\,s^{-1}}$, at $\eta=0.002$, $\Omega_b=32.5\,\mathrm{km\,s^{-1}\,kpc^{-1}}$, and $t=9.301\,\mathrm{Gyr}$. Opposite azimuthal sectors are combined: (a) $0^\circ$--$90^\circ$ and $180^\circ$--$270^\circ$; (b) $90^\circ$--$180^\circ$ and $270^\circ$--$360^\circ$. The relative prominence of positive- and negative-radial-velocity chevrons changes between the two sector pairs.}
    \label{fig:phase_space_chevrons} 
\end{figure}

Figure \ref{fig:phase_space_chevrons} shows the resultant phase-space plots of a Milky Way-like simulations with $\eta = 0.002$. The snapshot is at $t=9.301 \text{ Gyr}$ and has $\Omega_b = 32.5 \, \text{km s}^{-1} \text{kpc}^{-1}$. 

We column-normalise the data to show the fraction of matter in each radius bin (rather than a star count) to avoid biasing based on the underlying radial distribution of stars. The features are enhanced using an unsharp masking technique, whereby data is convolved with a Gaussian filter and then subtracting from a multiple of the original signal to make the chevrons more visible. 

The chevrons for the group with positive angular momentum are clearer and more prominent than those in the negative angular momentum dataset. Retrograde stars orbit in the opposite direction of the bar. Because they have a much larger relative velocity with the bar compared to the prograde population moving with the bar, the gravitational `tugs' do not compound as effectively, making physical resonant trapping much less efficient than for prograde stars. Also, the retrograde sample contains significantly fewer stars. These effects compound, causing the retrograde patterns to be much weaker than the prograde patterns. The $L_z>0$ group shall be the focus of our analysis. \label{ang mom preference}

The chevrons are asymmetric; at small radii Figure \ref{fig:chevrons_pos_45} shows a bias towards positive radial velocity whereas Figure \ref{fig:chevrons_pos_135} shows a bias towards negative radial velocity. This asymmetry switches sign in each quadrant of the galaxy. \label{asym}

These switches in asymmetry align with theoretical expectations for resonant orbits visualised by Figure 8 from the paper \citet{2024MNRAS.532.4389D} and described in Section \ref{math_framework}. Furthermore, consistent with the adiabatic sweeping framework, the chevron features naturally evolve over the galaxy's lifetime. As the bar decelerates, the resonances migrate outward, gradually capturing more stars and producing progressively stronger, larger-scale chevron signatures at higher radii.

\begin{figure}
    \centering
    \includegraphics[width=\linewidth]{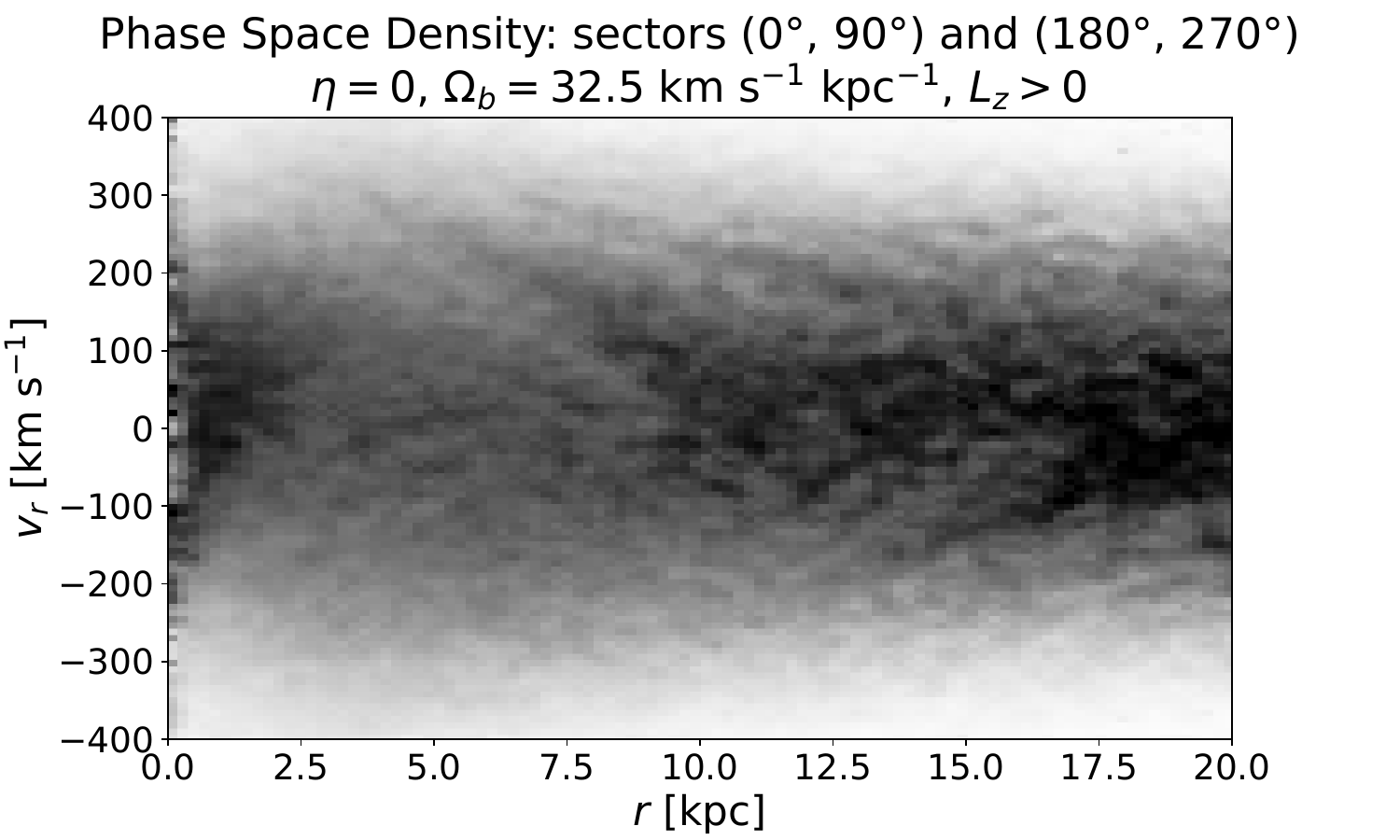}
        \caption{Simulated radial phase-space distribution of prograde particles ($L_z>0$) for a constant-pattern-speed Galactic bar, with $\eta=0$ and $\Omega_b=32.5\ {\rm km\,s^{-1}\,kpc^{-1}}$. The two opposite azimuthal sectors, $(0^\circ,90^\circ)$ and $(180^\circ,270^\circ)$, are combined. In contrast to the decelerating model shown in Figure \ref{fig:phase_space_chevrons}, the chevrons are substantially weaker and exhibit predominantly single-sided structures, reflecting the absence of resonant sweeping and the associated Euler-force-induced azimuthal displacement.}
    \label{fig:eta0chevrons} 
\end{figure}

For $\eta=0$ simulations, the chevrons are visible but are much fainter than in Figure \ref{fig:phase_space_chevrons}. The chevrons get slightly stronger throughout the lifetime of the simulation, but there is very little difference between early and late snapshots. An example is shown in Figure \ref{fig:eta0chevrons}.

The weakness of the signal is because `resonance sweeping' does not occur (because the bar does not slow) so stars are not collected over time in the same way. The small amount of strengthening over time happens because some stars happen to pass into resonance and get captured.

Another distinct difference between the constant speed case and the decelerating cases (Figure \ref{fig:phase_space_chevrons}) is in the morphology of their asymmetry. While the $\eta=0.002$ snapshots are highly asymmetric, they still exhibit weaker counterpart features across the $v_r=0$ axis. In contrast, the $\eta=0$ phase-space displays mostly single-sided chevrons; there is little discernible counterpart signal in the opposite velocity regime.

This is because when $\eta=0$ the asymmetry signal perfectly aligns with the bar axes (similar to the theoretical map in Figure 8 from \citet{2024MNRAS.532.4389D}). In the decelerating cases the Euler force causes the signal to shift azimuthally. As a result, part of the signal expected in the neighbouring quadrants is shifted into this one. This bleeding effect produces the other half of the chevron (but fainter). When there is no deceleration, there is no Euler force, so we only see the outflow section of the resonant orbit. This corresponds to only seeing stars with positive radial velocity. 

\subsubsection{Radial Asymmetry}

To further analyse this asymmetry in the phase-space chevron data (which is seen in Figure \ref{fig:phase_space_chevrons}) we quantify the asymmetry and its radial dependence. We calculate the total fractional asymmetry $A_r = \frac{N_+ - N_-}{N_+ + N_-}$ for each radial bin, where $N_+$ is the number of stars with positive radial velocity and $N_-$ is the number of stars with negative radial velocity.

The fractional asymmetry is chosen over the numerical asymmetry to prevent bias based on the number of stars at a fixed radius.

As expected the amount of fractional asymmetry varies radially. In Figure \ref{fig:chevrons_pos_45}, which contains data from the quadrants $(0^{\circ},90^{\circ})$ and $(180^{\circ},270^{\circ})$, in the inner region, the fractional asymmetry is positive but then decreases as radius increases.

\subsubsection{Spatial Asymmetry} \label{spatial method sims}

\begin{figure}
    \centering
    \includegraphics[width=\linewidth]{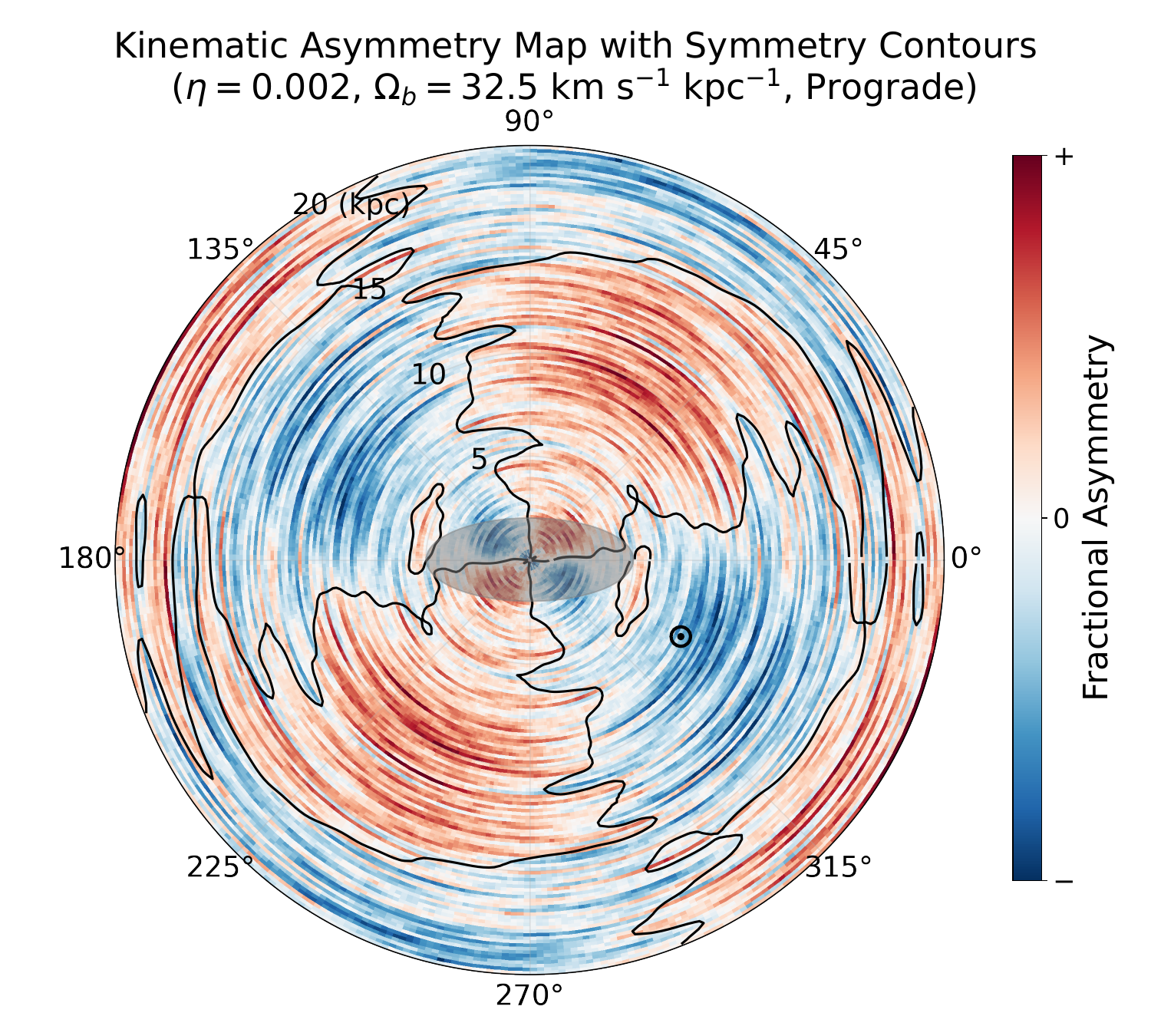}
        \caption{Spatial radial-velocity asymmetry of the simulated prograde population at $\eta=0.002$, $\Omega_b=32.5\,\mathrm{km\,s^{-1}\,kpc^{-1}}$, and $t=9.301\,\mathrm{Gyr}$, corresponding to Figure \ref{fig:phase_space_chevrons}. Red and blue indicate outward and inward excesses. Black contours derived from the Gaussian-smoothed map trace the asymmetry boundaries, and the grey ellipse marks the bar. The boundaries become azimuthally displaced from the bar axes as radius increases.}
    \label{fig:smoothed_contours_combined} 
\end{figure}

We extend this analysis to find the asymmetry as a function of both radius and azimuth to produce Figure \ref{fig:smoothed_contours_combined} as colour maps.

To maintain a sufficient signal-to-noise ratio across the phase-space maps, we compute the spatial asymmetry using overlapping azimuthal slices. Each slice is assigned a width of $30^\circ$, and the central angle is incrementally shifted by $1^{\circ}$ to map the full azimuthal range. Consequently, the `Fractional Asymmetry' represents an average over this $30^\circ$ window. This moving-average approach inherently smooths the data, suppressing the absolute peak magnitudes of the numerical asymmetry. The shape of the resonant orbits (specifically, the locations of the regions of inflow and outflow, and the boundaries between these regions) is the key information we are trying to extract, rather than the exact amplitude of the asymmetry itself. This spatial resolution is critical for tracing the resonances; furthermore, even with the dilution caused by azimuthal smoothing, the fractional asymmetry remains on the order of $0.01$, providing a robust signal for tracking the shapes of the resonant orbits. 

This calculation is very susceptible to noise at low radii, where the small spatial volume results in a very low star count. To remedy this, the extreme fractional asymmetry values are clipped at the $99.5$th percentile. This masking prevents artificial spikes from dominating the signal, effectively filtering out the innermost unreliable data.

We produce contours of symmetry by smoothing the data with a Gaussian filter, and then extracting the isocontours. This is then overlaid on the asymmetry plot.

Before quantifying $\Delta \phi_{\odot}$ (defined in Section \ref{math_framework}) caused by the Euler force, we first note the expected radial dependence on pattern speed. As the bar decelerates, the resonant orbits physically expand, causing the corresponding spatial asymmetry features to migrate outward to larger galactic radii. We observe this continuous outward expansion throughout our simulations, consistent with theoretical expectations. The specific present-day radii of these resonance boundaries were recently utilised by \citet{2024MNRAS.532.4389D} to calculate the pattern speed of the Milky Way's bar. 

With this radial behaviour confirmed, we now turn to the azimuthal misalignment. The asymmetry plot in Figure \ref{fig:smoothed_contours_combined} exhibits a clear azimuthal twist. In Figure 8 from \citet{2024MNRAS.532.4389D}, flips in asymmetry align with the principal axes of the bar. But in our Figure the flips are shifted at larger radius. This misalignment of asymmetry is a manifestation of the simulation having a decelerating bar. The simulated bar is decelerating at $\eta = 0.002$ and is therefore experiencing the Euler force. \label{noticing deceleration deviation}

\begin{figure*}
        \centering 
        \includegraphics[width=\linewidth]{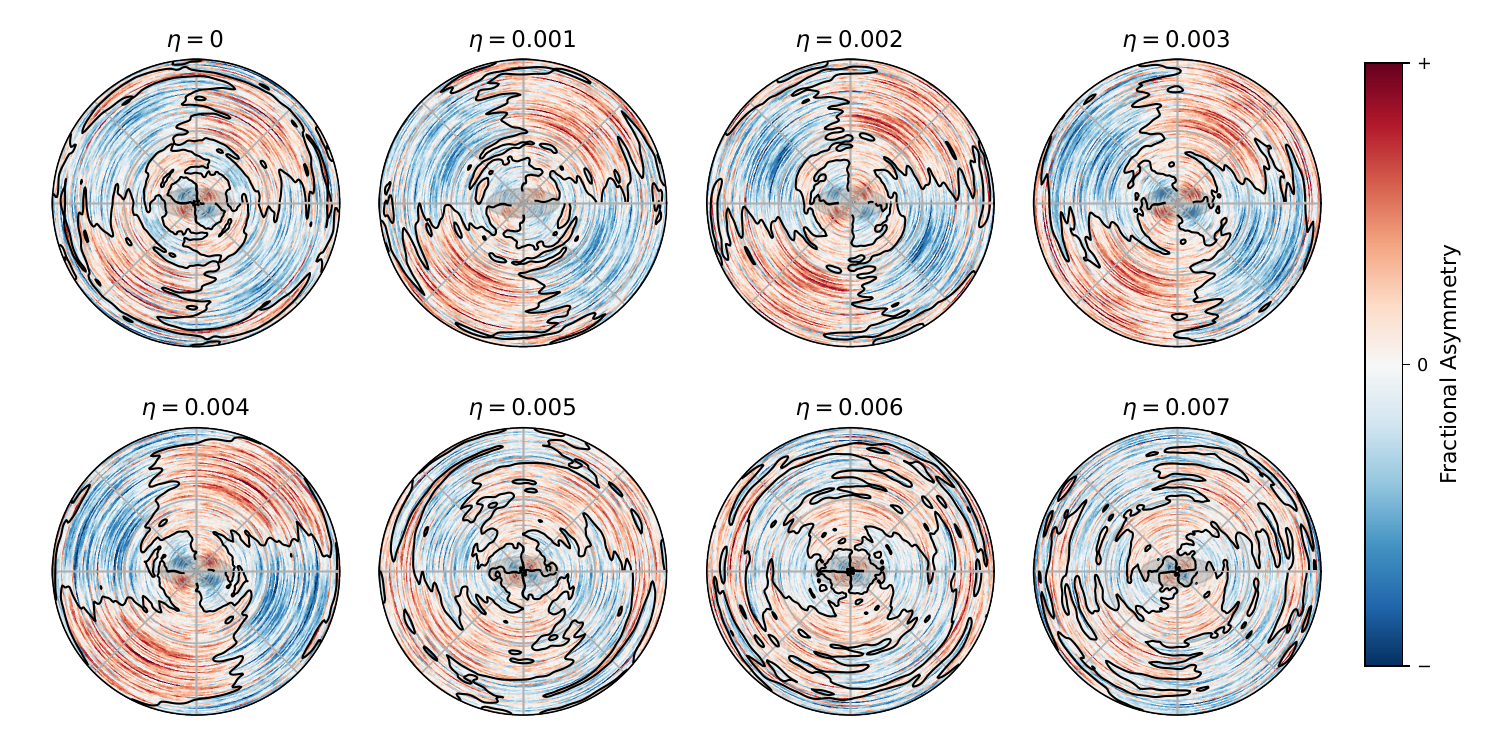}
        \caption{Simulated spatial radial-velocity asymmetry for prograde particles at a common pattern speed of $\Omega_b=32.5\,\mathrm{km\,s^{-1}\,kpc^{-1}}$, for the deceleration parameters indicated above each panel. Red and blue denote outward and inward excesses, and black curves show contours from the smoothed maps. The coherent azimuthal displacement is most evident at small positive $\eta$; the morphology becomes less regular around and above $\eta=0.004$ in these models. The $\eta=0$ panel is drawn from the constant-pattern-speed control simulation.}

        \label{fig:eta_grid_comparison}
\end{figure*}

$\Delta \phi_{\odot}$ (defined in Section \ref{math_framework}) increases with radius. So we ask the question: can we quantify this misalignment and does this effectively reflect the amount of deceleration? Figure \ref{fig:eta_grid_comparison} shows asymmetry snapshots at $\Omega_b = 32.5 \, \text{km s}^{-1} \text{kpc}^{-1}$ for a range of deceleration rates from $\eta = 0$ to $\eta=0.007$.

It is evident that $\Delta \phi_{\odot}$ increases as $\eta$ increases. The signal starts to break down between $\eta = 0.003$ and $\eta = 0.004$. This is likely because the simulations are no longer in the slow regime (discussed in Section \ref{math_framework}) leaving faint, irregular signatures. These high $\eta$ snapshots do not have the clear sloping deviation from the bar axes visible at smaller $\eta$. Instead of anti-clockwise twisting we see the signal sharply twist at small radii and then fold back on itself into messy clockwise arcs. In these systems the stars appear to not remain trapped, instead many are streaming out, leaving the observed signals. 

We focus our investigation on low deceleration systems between $\eta = 0$ and $\eta = 0.004$. These systems are more physically plausible for the Milky Way (as verified by the estimated $\eta$ values from \citet{Chiba2021} and \citet{Zhang_2025}) and they are more reliably quantifiable.

\subsection{Mock Catalogues} \label{mock}

\subsubsection{Mock Catalogue Generation} \label{mock catalogue}

To ensure a robust comparison between our idealised test-particle simulations and the real Milky Way, we model the observational uncertainties of the Gaia DR3 sample to create a mock catalogue. 

Uncertainties in stellar distances distort the spatial distribution inferred from observational data. Because a distance error displaces a star along the observer’s line of sight, structures appear elongated in the heliocentric radial direction, referred to as the `Finger of God' effect \citep{Jackson_1972}. In the context of this work, this manifests itself as a star's radius and azimuth being altered and therefore its radial velocity being incorrect (see also \citealt{Hey2023, Vislosky2024, Zhang2024}). This smearing effect can alter the apparent orientation of the asymmetry boundaries and therefore affect the phase measurement used to constrain bar deceleration.

To estimate the relevant uncertainty scale, we examine the fractional parallax uncertainties of the selected Gaia sample. Our quality cut requires $(|\varpi|/\sigma_\varpi>10)$, corresponding to fractional parallax uncertainties below 10$\%$, and the median fractional parallax uncertainty after all cuts is 6.6$\%$. We use these values as an approximate guide to the distance-error scale when constructing the mock catalogue. We adopt a $10\%$ fractional distance uncertainty for all particles, approximately the largest uncertainty permitted by our parallax cut. This is because the most uncertain stars, whose spatial displacements are largest, dominate the distortion of the asymmetry map. Our mock is therefore likely to overestimate the distance smearing (see Section \ref{discussion}).

To replicate this effect in our simulations, we transform each simulated particle into heliocentric observables (sky position, distance, proper motion and line-of-sight velocity), perturb the distance, and recompute Galactocentric positions and velocities with the proper motions and line-of-sight velocity held fixed. The angular-momentum selection is applied after this step, as for the Gaia data. We apply a Gaussian distance smear, injecting noise drawn from a normal distribution with a fractional error of $10\%$ ($\sigma = 0.10$). This injected noise scales linearly with the simulated star's distance from the observer. 

As described in Section \ref{phase space chevrons}, our analysis superimposes opposite sides of the galaxy as they are theoretically equivalent. To prevent superimposing the noisiest data (from the opposite side of the galaxy) on top of the cleaner data on the near side of the galaxy, we apply an azimuthal transformation before noise injection. We take all the data from the far side ($63^{\circ}<\phi<153^{\circ}$) and translate it $\phi \rightarrow \phi +180^{\circ}$. This mirrors the data to the near side, preserving the total particle count whilst avoiding the line-of-sight distance smearing that would have dominated the far-side signal. We then add the distance noise to the data. The analysis then naturally fills in the gap at the other side of the galaxy. The result is the data at $\phi$ and $\phi +180^{\circ}$ is equivalent.

We also apply a Gaia-like distribution cut. However, because the Gaia-matched mock retains at most one simulated particle per Gaia star (Appendix \ref{gaia fitting}), it contains $\lesssim 3 \times 10^5$ particles, too few for a reliable phase measurement in many snapshots. We therefore use the full distance-smeared mock as our baseline calibration. The results for this experiment can be found in Appendix \ref{gaia fitting}.

\subsubsection{Success of the Catalogue}

\begin{figure}
    \centering
    \includegraphics[width=\linewidth]{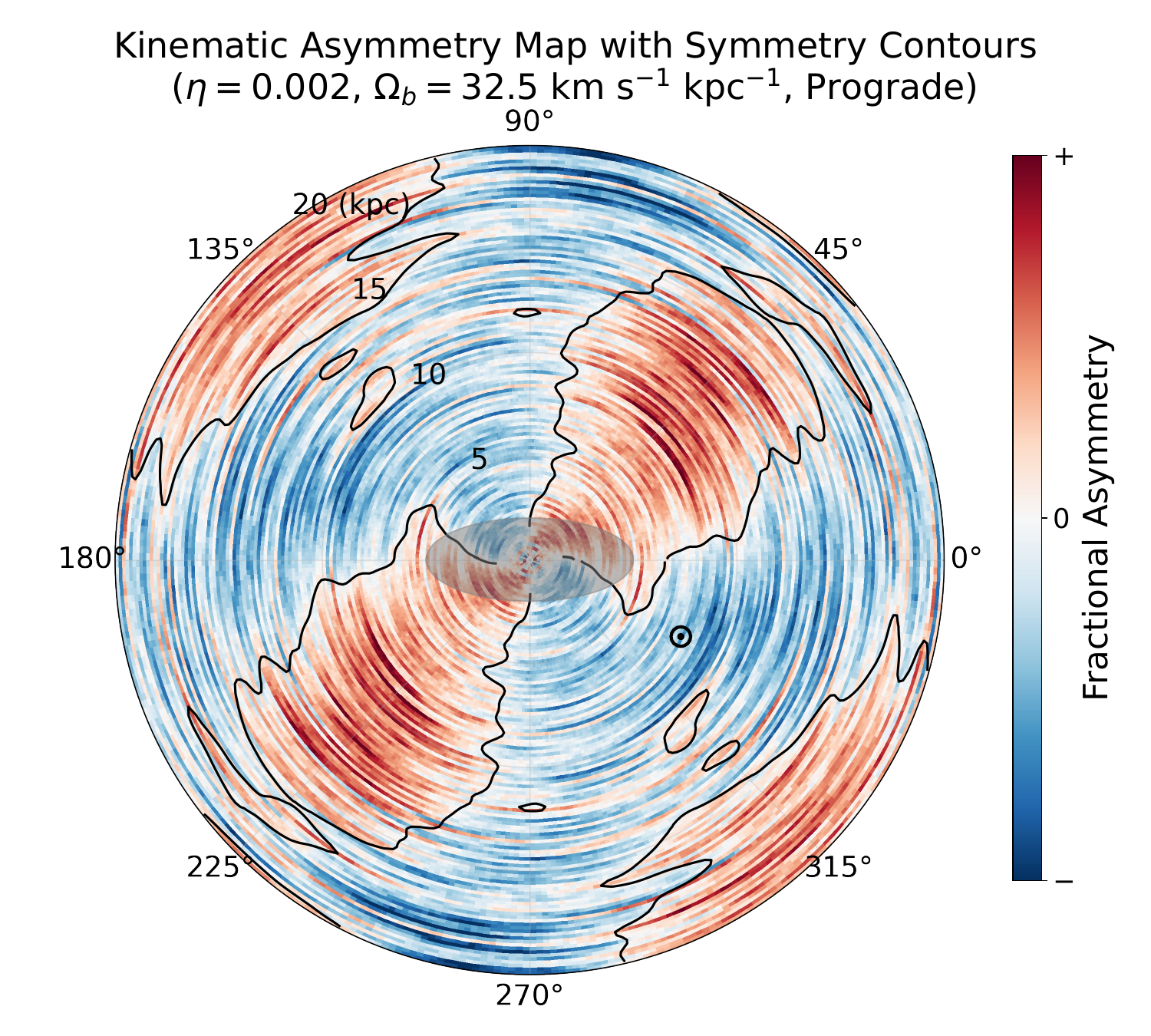}
        \caption{Spatial radial-velocity asymmetry of the mock prograde stellar population at $\eta=0.002$ and $\Omega_b=32.5\ {\rm km\,s^{-1}\,kpc^{-1}}$, after the injection of heliocentric distance uncertainties. Red and blue indicate excesses of outward- and inward-moving stars, respectively, while black contours trace the boundaries of the smoothed asymmetry map. The radial elongation of the features illustrates the artificial Finger-of-God effect introduced by distance smearing. Despite this distortion, the large-scale azimuthal twist associated with the decelerating bar remains visible.}
    \label{fig:smeared} 
\end{figure}

An example of the spatial asymmetry map of the mock catalogue for the $\eta=0.002$, $\Omega_b=32.5$ snapshot is presented in Figure \ref{fig:smeared}.

There is the expected artificial `Finger of God' radial elongation. Crucially, however, the large-scale azimuthal twist induced by the Euler force survives these stringent observational constraints and remains resolvable, although altered.

\subsection{Measuring Deviation} \label{mcmc_fitting_section}

To derive a simple observational metric for the Euler force, we quantify the azimuthal misalignment between the bar axes and the asymmetry signal. We measure $\Delta \phi_{\odot}$ (defined in Section \ref{math_framework}) across the annulus $R_{\odot} \pm 1.25 \, \text{kpc}$ to capture the highest volume of Gaia DR3 data (which is centred about the Sun). This angle is a consequence of the system decelerating so will be a direct probe for the magnitude of deceleration.

Summing the fractional asymmetry over this annulus would yield an approximately sinusoidal profile as a function of azimuth. The peaks and troughs of this curve directly correspond to the different regions of positive and negative fractional asymmetry seen in the spatial maps (Figure \ref{fig:smoothed_contours_combined}). Crucially, because the raw spatial data can be highly complex and noisy, fitting a sinusoidal curve to this profile provides an averaged representation of the asymmetry. 

The fit is found using an Affine-Invariant Markov Chain Monte Carlo (MCMC) ensemble sampler, which optimises the parameters. We fit the fractional asymmetry data to a sinusoid:

\begin{equation} \label{eq:sine_fit}
    Y(\phi) = A \sin(2(\phi - \phi_0))
\end{equation}

where $Y(\phi)$ is the fractional asymmetry at azimuth $\phi$, $A$ is the amplitude of the $m=2$ mode, and $\phi_0$ is the phase parameter which corresponds to the azimuth of the zero crossing.

To handle the measurement uncertainties without artificially scaling or flattening the empirical error bars, the MCMC evaluates the fit using a complete Gaussian log-likelihood. 

The total variance for each azimuthal bin ($\sigma_{\text{total}, i}^2$) is modelled by adding the physical bin measurement error ($\epsilon_i$) in quadrature with an MCMC-fitted intrinsic scatter parameter ($\sigma$):

\begin{equation} \label{eq:total_variance}
    \sigma_{\text{total}, i}^2 = \sigma^2 + \epsilon_i^2 
\end{equation}
  
This intrinsic variance parameter acts as a statistical counterweight, allowing the solver to accommodate localised regions of underestimated noise without mathematically blowing up the likelihood or flattening the physical amplitude of the signal. The MCMC optimises the fit by evaluating the squared residuals (the $\chi^2$ statistic) relative to this total variance via the Gaussian log-likelihood ($\ln \mathcal{L}$):

\begin{equation} \label{eq:log_likelihood}
    \ln \mathcal{L} = -\frac{1}{2} \sum_{i} w_i \left[ \frac{(Y_i - Y(\phi_i))^2}{\sigma_{\text{total}, i}^2} + \ln(2\pi\sigma_{\text{total}, i}^2) \right]
\end{equation}

where $w_i$ is a spatial weighting mask, defined as such:

\begin{equation}
    w_i = \left(\frac{1 + \text{cos}(2x_{valid} - 333\pi/90)}{2} \right)^4 + e^{-5}
\end{equation}

where $x_{valid}$ is in radians. The fit is therefore dominated by the azimuths least affected by distance errors, and largely ignores the artificial `Finger of God' signal far from the Sun.

We determine $\Delta \phi_{\odot}$ from the MCMC posterior distribution of $\phi_0$.

In the mock data the distance uncertainty means there is an artificial change to the signal. To prevent the fit from being skewed by dominant, artificial structural peaks located far from the reliable data, weighting is applied to favour the data around Solar radius. This spatial weight is applied directly to the log-likelihood evaluations. The mask utilises a steep, exponentiated cosine taper centred exactly on the target axes ($333^\circ$ and $153^\circ$). This means that the fitting function largely ignores the artificial `Finger of God' signal at the furthest point from the Sun. Red lines mark the Solar azimuth and its folded counterpart on the other side. 

To encapsulate uncertainties in the length of the Milky Way's bar \citep{Hilmi2020} and therefore the length of the simulated bars, an additional error $\sigma^2_{\text{bar}}$ is included, which varies with $\eta$ (see Appendix \ref{appendix_bar_length} for a detailed sensitivity analysis). After the measurement uncertainty for each snapshot is found, this bar length error is added to it in quadrature.

\begin{figure*}
    \centering
    \begin{subfigure}[b]{\columnwidth}
        \centering
        \includegraphics[width=\linewidth]{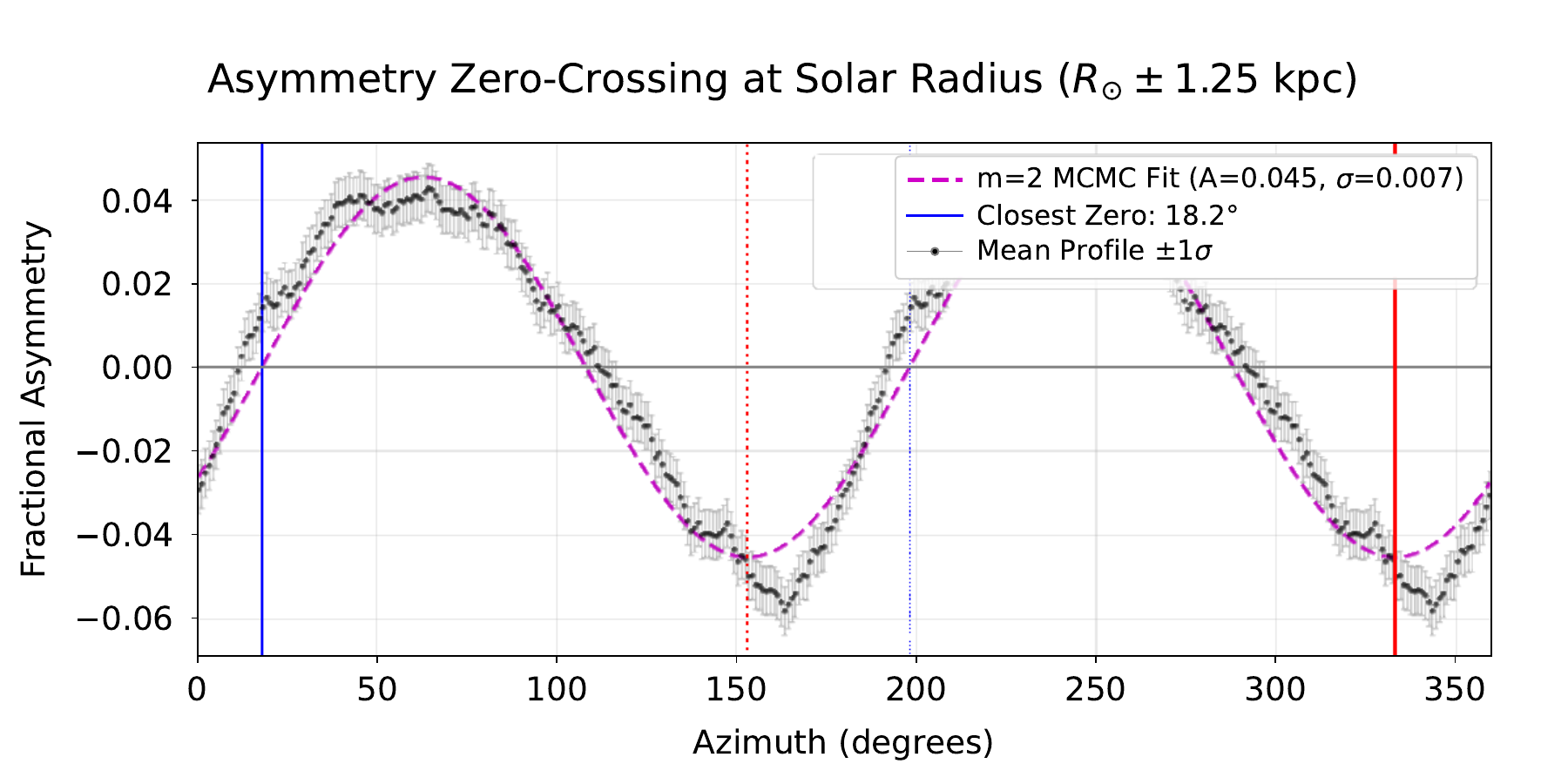}
        \caption{Deviation sinusoid for the raw simulated data.}
        \label{fig:dev}
    \end{subfigure}
    \hfill
    \begin{subfigure}[b]{\columnwidth}
        \centering
        \includegraphics[width=\linewidth]{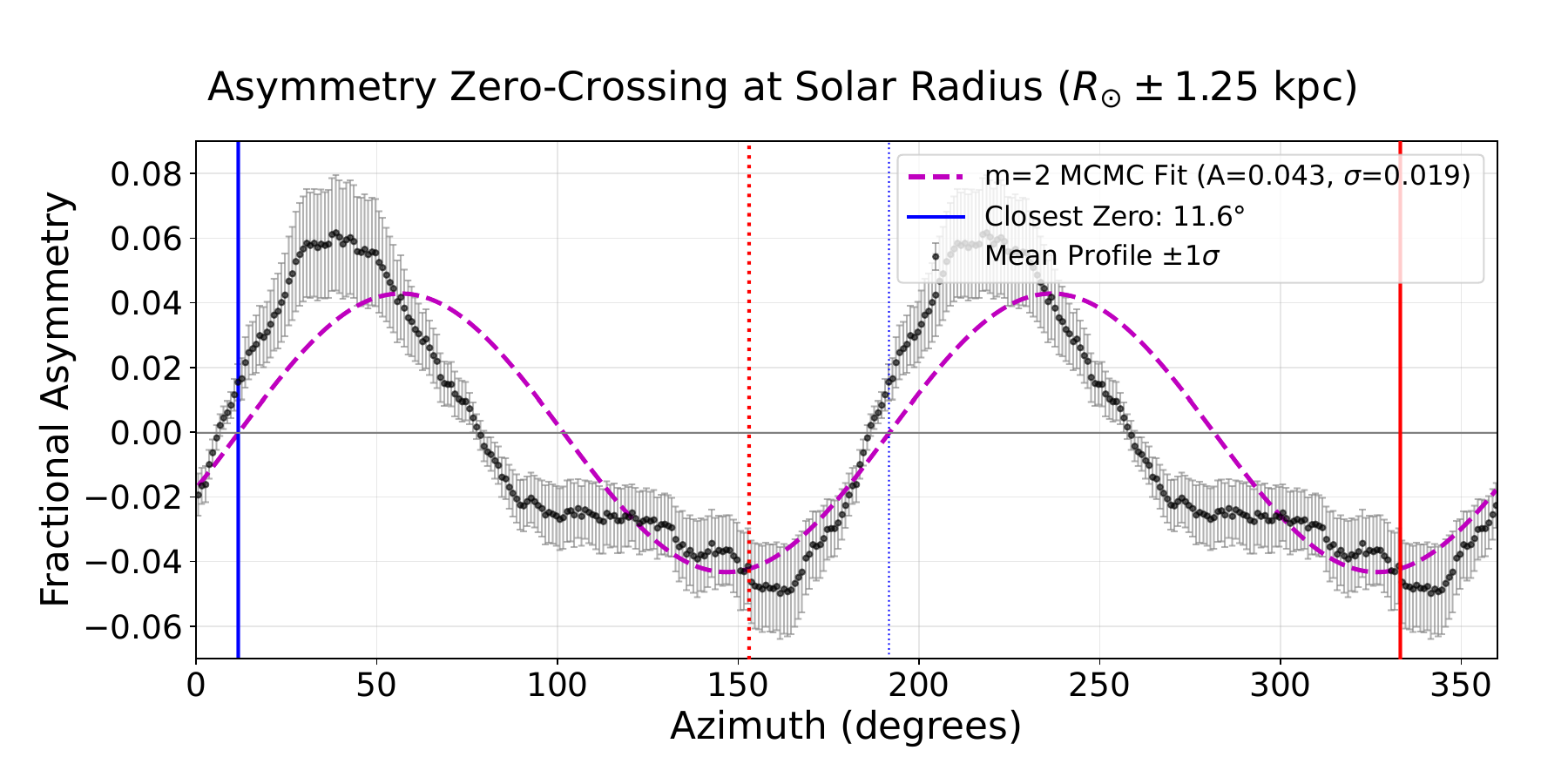}
        \caption{Deviation sinusoid for the noise-injected simulated data.}
        \label{fig:dev noisy}
    \end{subfigure}
    \vspace{0.5cm}
    \begin{subfigure}[b]{\columnwidth}
        \centering
        \includegraphics[width=\linewidth]{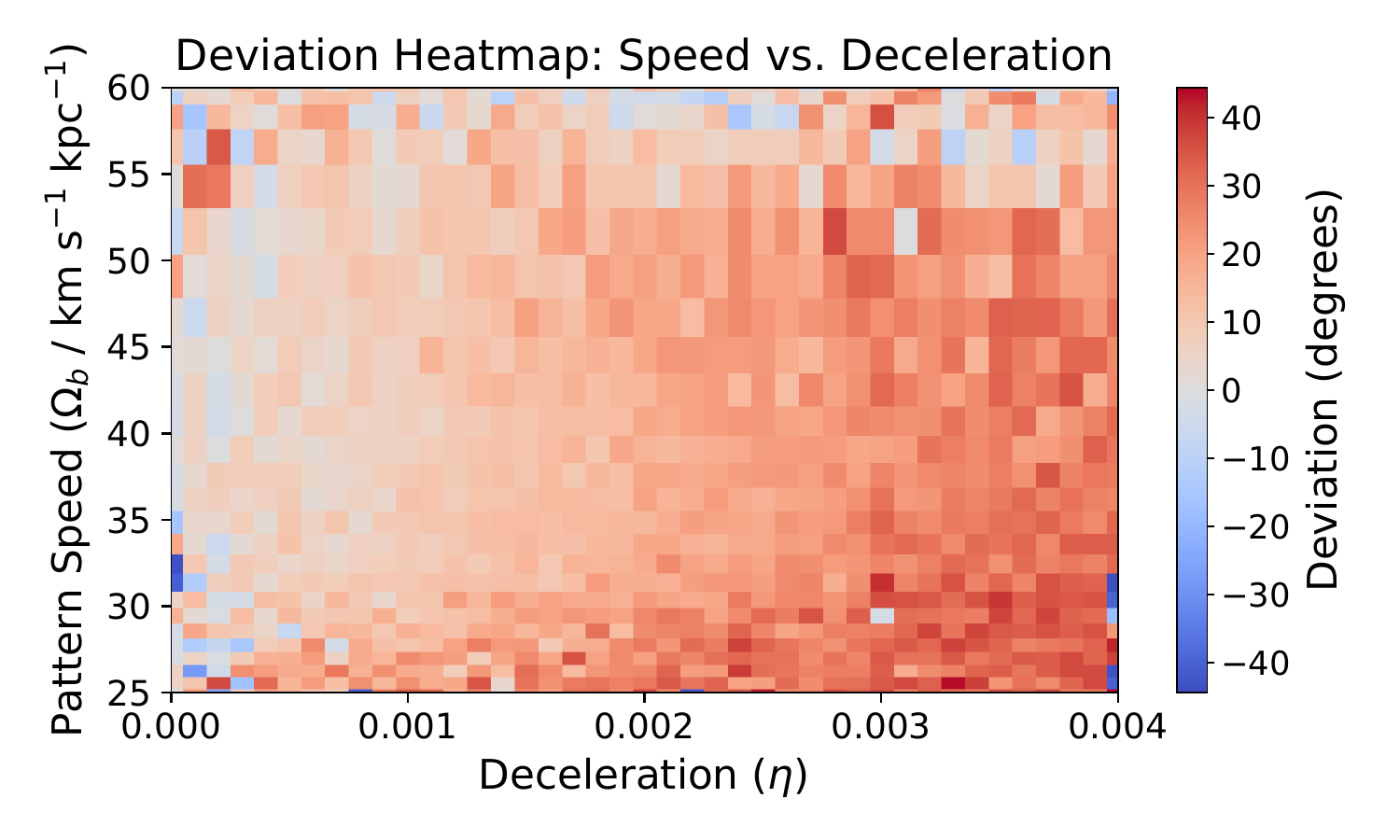}
        \caption{Deviation heatmap for the raw simulated data.}
        \label{fig:heatmap}
    \end{subfigure}
    \hfill
    \begin{subfigure}[b]{\columnwidth}
        \centering
        \includegraphics[width=\linewidth]{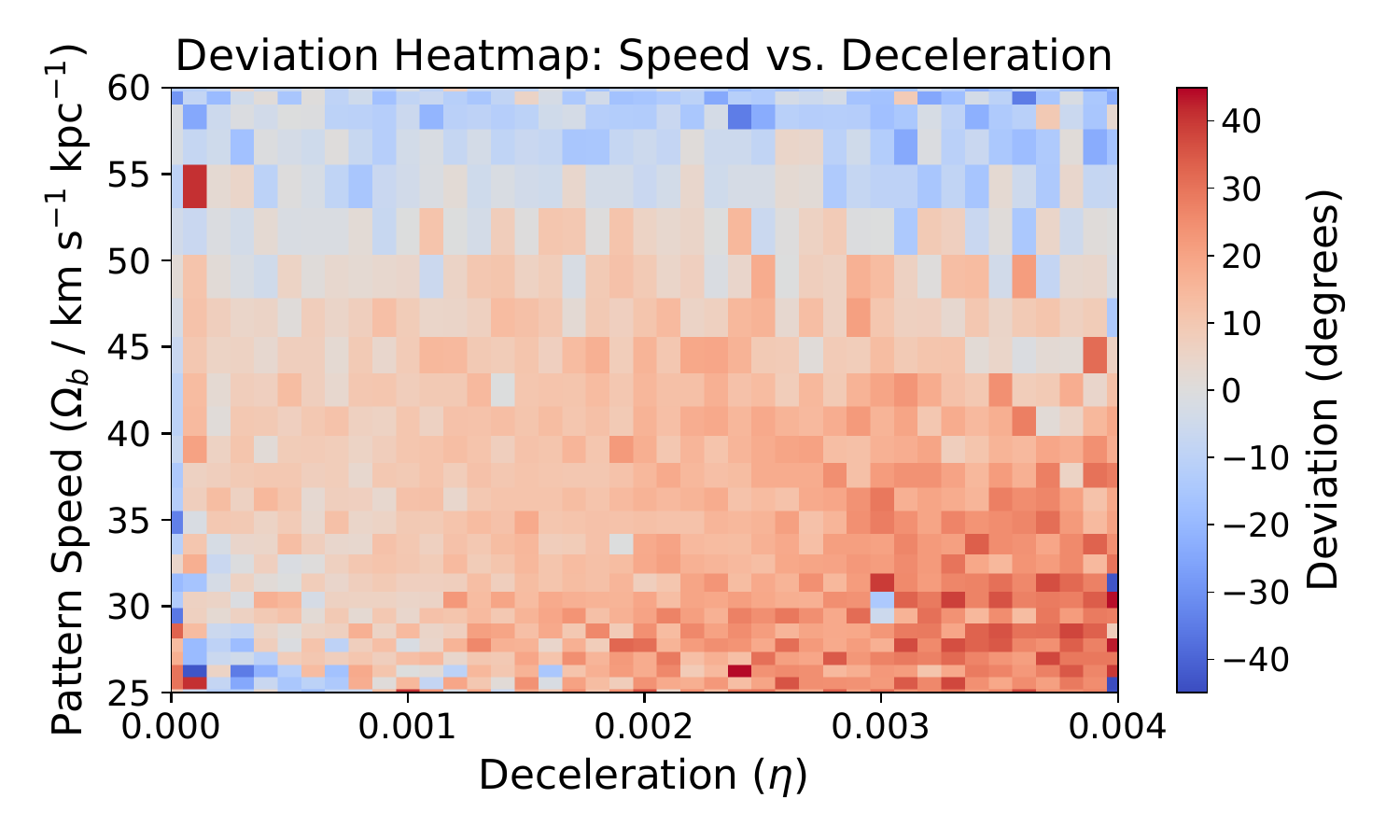}
        \caption{Deviation heatmap for the noise-injected simulated data.}
        \label{fig:heatmap_noisy}
    \end{subfigure} 

    \caption{Measurement of the azimuthal phase of radial-velocity asymmetry in the simulated prograde population, before and after heliocentric distance errors are applied. Panels (a) and (b) show example profiles in the annulus $R_\odot\pm1.25\,\mathrm{kpc}$ for the idealised and distance-smeared simulations, respectively. Black points and error bars show the measured profiles, magenta dashed curves show the fitted $m=2$ sinusoids, blue lines mark the selected zero crossings and their 180-degree counterparts, and red lines mark the Solar azimuth and its folded counterpart. Panels (c) and (d) show the inferred $\Delta \phi_{\odot}$, in degrees, across the $\eta$--pattern-speed grid for the idealised and distance-smeared simulations, respectively. Distance errors alter the recovered phase and its dependence on pattern speed; the smeared profiles are fitted with the angular weighting described in Section \ref{mcmc_fitting_section}.}
    \label{all} 
\end{figure*}

For the raw simulation data there is no distance uncertainty so no weighting is applied, an example can be found in Figure \ref{fig:dev}. An example from the mock catalogue with the weighting applied can be seen in Figure \ref{fig:dev noisy}.
 
We perform this analysis for our simulated data with and without distance smearing across all of speed-deceleration space, to produce Figure \ref{all}. 

The $m=2$ sinusoidal model seen in Figure \ref{fig:dev} provides a robust fit to the data. While localised deviations from a pure sinusoid exist, such as near $\phi \approx 160^{\circ}$ and $340^{\circ}$, these are naturally incorporated into the intrinsic scatter parameter ($\sigma$). Crucially, this approach accurately finds the zero-crossing azimuth, our primary metric for the deceleration shift.

Varying $\eta$ and $\Omega_b$ generally produces well-behaved sinusoidal profiles with phase shifts scaling with the deceleration rate. However, the fitting routine naturally degrades in a few specific kinematic regimes. 

First, at certain parameter combinations (such as $\eta=0$ at low $\Omega_b$), the Solar annulus aligns directly with a transition boundary between two distinct resonances. In these boundaries, the asymmetry profile captures a complicated superposition of both resonances, resulting in a flat, unconstrainable signal. A detailed investigation into the effect of these transition zones is provided in Appendix \ref{transition zones}.

Second, the fits become inherently noisier in some regions of speed-deceleration space. For example, at high pattern speeds ($\Omega_b \gtrsim 50 \, \text{km s}^{-1} \text{kpc}^{-1}$) we see more noise because these snapshots correspond to early times in the simulation, when resonant sweeping has not yet accumulated enough stars to produce a strong macroscopic signal. Furthermore, the resonant features are physically narrower at high pattern speeds, increasing the likelihood of transition zones crossing the Solar annulus. Both effects diminish the signal-to-noise ratio. 

Plotting $\Delta \phi_{\odot}$ across speed-deceleration space yields Figure \ref{fig:heatmap}. There are some points where the signal-to-noise ratio is too low; this is marked by a cross. This Figure clearly shows an increase from $\Delta \phi_{\odot}=0$ to large $\Delta \phi_{\odot}$ as deceleration increases. As pattern speed is varied there does not seem to be a clear trend other than varying noise. This is the result that we have been working towards; this calibration quantifies how $\Delta \phi_{\odot}$ depends on the deceleration of the bar. It relies, however, on idealised data with perfect distances and complete spatial coverage; we therefore repeat the measurement on the mock catalogue.

Figure \ref{fig:dev noisy} shows the fit traces the data more closely around $\phi = 153^{\circ}$ and $\phi = 333^{\circ}$, where the data has been least affected by observational effects. It does still take into account the rest of the curve but does not fit as closely.

Performing the weighted fitting routine across all of the noisy simulation data produces the heatmap in Figure \ref{fig:heatmap_noisy}.

Information has been lost between Figure \ref{fig:heatmap} and Figure \ref{fig:heatmap_noisy}. At low $\eta$ there is some trend of low to high $\Delta \phi_{\odot}$ as hoped, however all of the data points have been shifted to a significantly lower $\Delta \phi_{\odot}$. This is averaging between the true value (some angle larger than $0^{\circ}$) and the azimuth at which there is no `Finger of God' effect (the azimuth of the Sun which is at $-27^{\circ}$). 

\begin{figure}
    \centering
    \includegraphics[width=\linewidth]{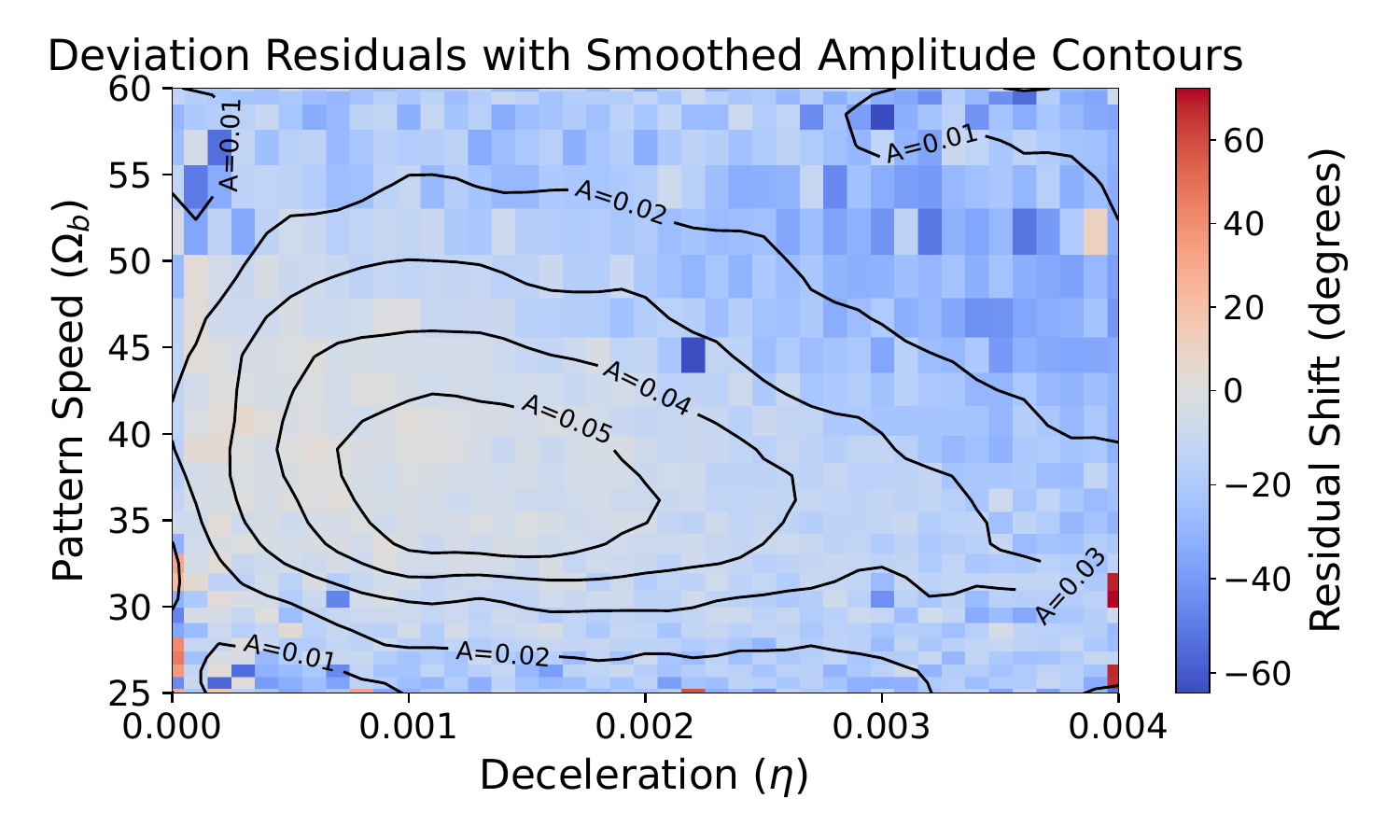}
        \caption{Change in the recovered asymmetry phase after heliocentric distance errors are applied, plotted across the simulation grid. The residual is $\Delta \phi_{\odot}$ in the distance-smeared snapshot minus $\Delta \phi_{\odot}$ in the idealised snapshot, using the same angular branch convention as Figure \ref{all}. Negative values indicate a reduction in recovered $\Delta \phi_{\odot}$. The shift varies across both $\eta$ and pattern speed, demonstrating the need for an observational calibration. The contours show the fitted $m=2$ amplitude of the radial-velocity fractional-asymmetry profile in the Solar annulus across the $\eta$--pattern-speed grid. The amplitude indicates where the azimuthal phase signature is strong in the adopted simulations.}
        \label{fig:amplitude_heatmap}
\end{figure}

The difference between the crossing value before and after distance noise has been added is shown in Figure \ref{fig:amplitude_heatmap}.

There is some variation with $\eta$ and $\Omega_b$ in the residuals, as shown by the background heatmap in Figure \ref{fig:amplitude_heatmap}. This variation is a direct consequence of the strength of the raw asymmetry signal. When the asymmetry signal is strong the `Finger of God' effect has minimal effect on what we measure, however if the raw asymmetry signal is weaker the artificial distance noise dominates and the residuals are larger. The effect of the distance errors depends on pattern speed, so we must take pattern speed into account when inferring the deceleration parameter. Previously $\Delta \phi_{\odot}$ was purely a probe of the Euler force and therefore the deceleration of the bar - now pattern speed must be taken into account due to this observational artifact.

This connection between the residuals and the strength of the raw asymmetry signal is evidenced by the alignment of the contours with the heatmap. The contours show the strength of the asymmetry signal (measured by the amplitude of the fit for the asymmetry sinusoids before smearing, i.e. $A$ in Figure \ref{fig:dev}). In Figure \ref{fig:amplitude_heatmap} the contours showing asymmetry strength align very closely with the heatmap showing the residuals between the $\Delta \phi_{\odot}$ measurements before and after distance smearing.
 
Figure \ref{fig:eta_grid_comparison} shows that the signal is weak at $\eta = 0$ since no resonance sweeping occurs, but otherwise the signal generally gets weaker as $\eta$ increases. 

The other trend we see is that the true signal is much weaker at higher pattern speeds so the resultant angle is much closer to the `Finger of God' zero point. A quantification of the signal strength across speed-deceleration space is shown in Figure \ref{fig:amplitude_heatmap}. Similarly, the distribution of the errors are shown in Figures \ref{fig:errors} and \ref{fig:errors noisy} in Appendix \ref{strength}.

\section{Observational Data} \label{methods}

\subsection{Data Preparation} \label{gaia methods}

To replicate and extend the analysis of the phase-space chevrons first located in \citet{2023MNRAS.518.6200B} in the Gaia Data Release 3 (DR3) dataset \citep{2021A&A...649A...1G} we isolate a halo sample of the stellar population interacting with the bar's resonances. We apply a set of kinematic and spatial filters and a set of quality cuts. 

First, we calculate the vertical component of the angular momentum ($L_z$) for each star. We adopt a Solar position of $r= 8.122 \, \text{kpc}$ \citep{GRAVITY:2018ofz} and $(v_x,v_y,v_z)=(12.9, \, 245.6, \, 7.78) \text{ km s}^{-1}$ \citep{Schonrich2010, 2018RNAAS...2..210D}. We apply a kinematic cut, restricting the sample to stars with $|L_z| < 500 \, \text{kpc km s}^{-1}$. This excludes stars on disc-like orbits and bounds our analysis to the relevant phase-space regions. Additionally, a geometric distance mask is applied to retain only stars within $15\, \text{kpc}$ of the Sun. We also require a distance signal-to-noise ratio ($D/\sigma_D$) strictly greater than $3$. To ensure high-quality astrometry, we enforce a parallax signal-to-noise cut of $|\varpi|/\sigma_\varpi > 10.0$. Finally, an upper reddening limit of $E(B-V) < 100 \text{ mmag}$ is applied. 

Then we transform the Gaia phase-space coordinates into a bar-aligned reference frame. In standard Galactic coordinates (where rotation is clockwise), the Sun is located at an azimuth of $180^\circ$. We flip the Gaia dataset to change the direction of rotation and then apply a rotation of $-153^\circ$ in order to match the simulated data for a direct comparison. This transformation results in a solar azimuth of $333^{\circ}$. \label{transformation}

The initial dataset contains $33,581,727$ stars, which reduces to a final sample of $292,504$ stars after all cuts are applied. This filtered sample is sufficient to resolve the subtle phase-space features of the bar's resonances.

\subsection{Recovering Phase-Space Chevrons} \label{gaiachevrons}

We first reproduce the phase-space chevrons \citep{2023MNRAS.518.6200B}. The method for extracting phase-space chevrons from the Gaia data mirrors the methodology used to isolate the phase-space chevrons in the simulated data shown in Figure \ref{fig:phase_space_chevrons}. The data preparation results in the Sun being positioned at $r=8.122 \text{ kpc}$ \citep{GRAVITY:2018ofz} and $\phi=333^{\circ}$ \citep{Wegg_2015, 2019MNRAS.490.4740B}, where the system rotates anti-clockwise; this matches the processed simulated data. The data from opposite Galactic quadrants is superimposed to maximise the signal-to-noise ratio. 

In the Gaia data most of the stars in the filtered dataset are in the quadrant containing the Sun, there is not enough data to effectively produce the phase-space chevrons, so for now we only see the quadrants $(90^{\circ},180^{\circ})$ and $(270^{\circ},360^{\circ})$.

We column-normalise and enhance the data using an unsharp masking technique, like in Section \ref{phase space chevrons}. 

\begin{figure*}
    \centering
    \includegraphics[width=\linewidth]{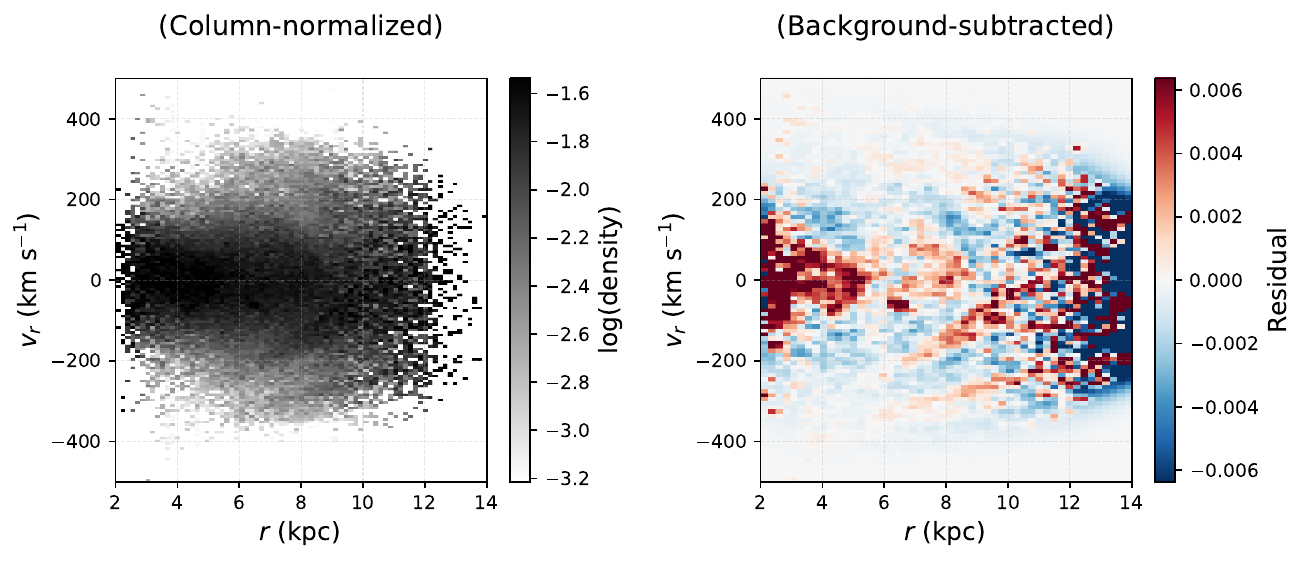}
    
    \caption{Radial phase-space distribution of the prograde Gaia DR3 sample selected in Section \ref{gaia methods}. Left: logarithm of the column-normalised stellar distribution, with each radial bin normalised over radial velocity. Right: residual after subtraction of a smoothed background, highlighting the chevron-like substructure. The sample satisfies $0<L_z<500\,\mathrm{kpc\,km\,s^{-1}}$ and is shown in the bar-aligned frame defined in Section \ref{gaia methods}.}

    \label{fig:gaia_recovered_chevrons_2panel}
\end{figure*}

Figure \ref{fig:gaia_recovered_chevrons_2panel} presents the phase-space distributions of the prograde ($L_z > 0$) stellar population. The left panel displays the column-normalised stellar density, while the right panel presents the corresponding unsharp-masked residuals.

The prograde stellar population displays distinct chevron structures that are asymmetric about $(v_r=0)$. This asymmetry suggests that they are a dynamical consequence of the Galactic bar. As merger debris phase-mixes, its distribution in radial phase space becomes increasingly symmetric between inward and outward moving stars. \citet{2024MNRAS.531.1422D} demonstrates this evolution in simulations and find that the observed asymmetry favours dynamically young debris over the phase-mixed remnants of an ancient merger. The asymmetry in our observational sample therefore challenges an interpretation based solely on ancient, phase-mixed merger debris, although it does not by itself eliminate a recent merger from a response to the Galactic bar. Bar-resonance models provide an alternative explanation, predicting an asymmetry whose sign varies with Galactic radius and azimuth \citep{2024MNRAS.532.4389D}. 

We therefore examine how the asymmetry varies with Galactic position and how its boundaries align with the bar.

\subsection{Spatial Asymmetry} \label{spatial method}

\begin{figure}
        \centering
        \includegraphics[width=\linewidth]{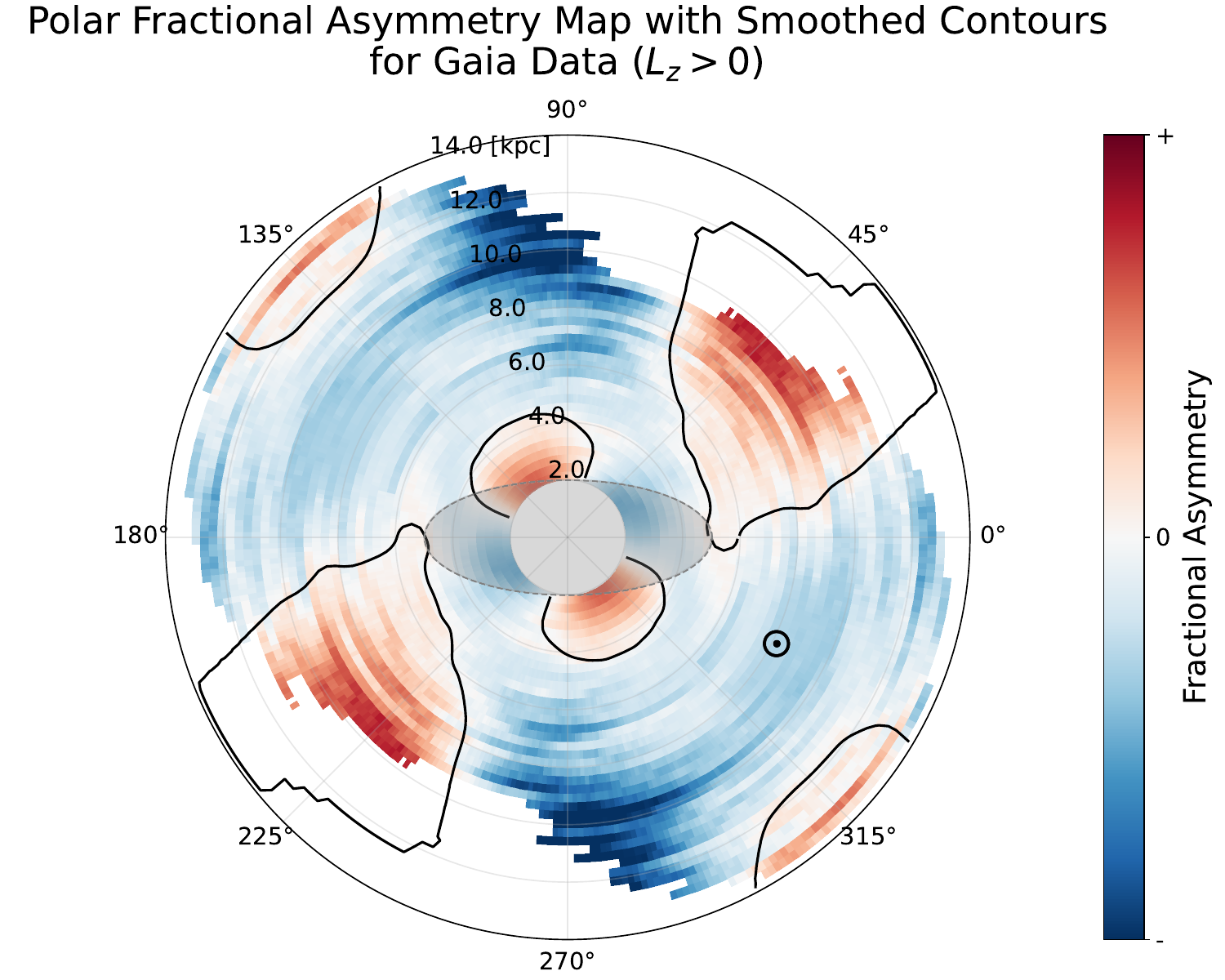}
    \caption{Spatial distribution of radial-velocity fractional asymmetry in the prograde Gaia DR3 sample, shown in the Galactic plane in the bar-aligned frame. Red and blue indicate an excess of outward- and inward-moving stars, respectively, using $A_r=(N_+-N_-)/(N_++N_-)$. The asymmetry is evaluated in overlapping 30-degree azimuthal windows advanced by 1 degree; black curves show contours derived from the smoothed map. The grey ellipse marks the bar orientation. Both the intrinsic kinematic structure and heliocentric distance errors can contribute to the apparent orientation of the boundaries.}
    \label{fig:gaia_asym_flips_pos}
\end{figure}

We generate a smoothed spatial asymmetry map of the Gaia dataset, this is presented in Figure \ref{fig:gaia_asym_flips_pos}, using the overlapping $30^{\circ}$ azimuthal slices described in Section \ref{spatial method sims}. 

The map reveals clearly defined transition boundaries between regions of kinematic inflow and outflow. Importantly, these transition boundaries are not perfectly aligned with the axes of the Galactic bar. This kinematic signature extracted from resonant orbits seems to show the expected Euler twisting. However, we appear to be observing this through the lens of distance smearing. At extreme distances from the observer (the Sun's position of $r = 8.122 \text{ kpc}$ and $\phi = 333^{\circ}$), the features align with the azimuth of the Sun, suggesting these are primarily distance error signals. 

If the observed asymmetry were solely an artifact of distance uncertainties, the signal would be a strictly antisymmetric signal, mirrored across the Solar azimuth. Instead, Figure \ref{fig:gaia_asym_flips_pos} breaks this symmetry, displaying a broad, continuous region of negative asymmetry (indicated in blue) surrounding the position of the Sun. This suggests that the spatial asymmetry map is some superposition of the true resonant orbit signal and the expected line-of-sight observational smearing. This is analogous with our mock dataset generated in Section \ref{mock}.

\subsection{Deviation from the Bar Axes} \label{fin_dev}

To extract $\Delta \phi_{\odot}$ (defined in Section \ref{math_framework}) for the Gaia data we apply the same method as described in Section \ref{mcmc_fitting_section}. Following Section \ref{mcmc_fitting_section}, we extract the asymmetry profile in the annulus \(R_\odot\pm1.25\,\mathrm{kpc}\). This produces an approximately sinusoidal profile as a function of azimuth, but with artificial peaks due to the finger-of-God effect. The peaks and troughs of this curve directly correspond to the alternating quadrants of positive and negative fractional asymmetry seen in the spatial map (Figure \ref{fig:gaia_asym_flips_pos}).

\begin{figure}
        \centering
        \includegraphics[width=\linewidth]{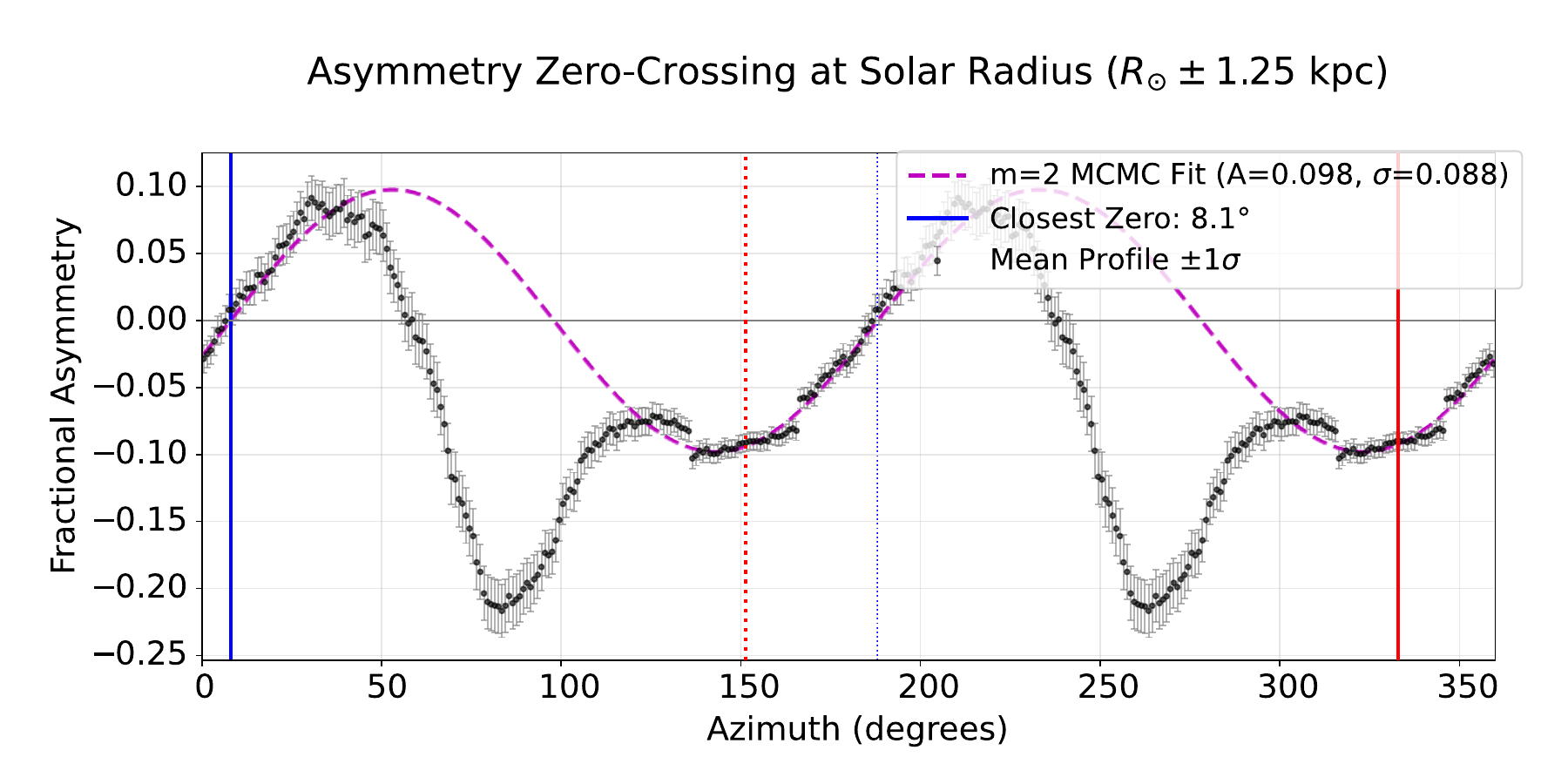}
        \caption{Azimuthal profile of radial-velocity fractional asymmetry for the prograde Gaia DR3 sample in the annulus $R_\odot\pm1.25\,\mathrm{kpc}$. Black points with error bars show the measured profile; the magenta dashed curve is the fitted $m=2$ sinusoid. Blue vertical lines mark the selected fitted zero crossing and its 180-degree counterpart, while red lines mark the Solar azimuth and its folded counterpart. The displayed crossing is approximately 8.1 degrees from the bar major axis. The fit includes intrinsic scatter and the angular weighting described in Section \ref{mcmc_fitting_section}.}
        \label{fig:dev_at_Sun_r_gaia}
\end{figure}

Figure \ref{fig:dev_at_Sun_r_gaia} presents the sinusoidal fit to the observational asymmetry data.

We extract the phase offset directly from the resulting posterior distributions. From this, we measure $\Delta \phi_{\odot} = 8.06^\circ \pm 2.36^\circ$ for the Gaia observational data. 

To ensure our $\Delta \phi_{\odot}$ measurement takes into account structural uncertainties in the Galactic potential, we include a systematic error of $\sigma_{\text{bar angle}} = 2^{\circ}$ to account for uncertainty in the bar angle. This error is added to the fit error to produce a final Gaia measurement of $\Delta \phi_{\odot} = 8.06^\circ \pm 3.09^\circ$.

\section{Measuring Deceleration} \label{observation results}

In this section, we compare the key results obtained from processing the Gaia DR3 data and our simulated models. 

In Section \ref{spatial method sims} we map the spatial distribution of these features. In Section \ref{mock} we create a mock catalogue of simulations and apply the same analysis for each snapshot. Then in Section \ref{mcmc_fitting_section} we quantify $\Delta \phi_{\odot}$ (defined in Section \ref{math_framework}) in each case. As shown in Figure \ref{fig:gaia_recovered_chevrons_2panel}, we successfully recover the phase-space chevrons previously identified by \citet{2023MNRAS.518.6200B} and \citet{2024MNRAS.532.4389D}, and then extend this analysis in Sections \ref{spatial method} and \ref{fin_dev}, to get an analogous measurement of $\Delta \phi_{\odot}$ for the observed data. Now we make the comparison.

The formal error does not include systematic uncertainties from distance errors. These are accounted for, to first order, by calibrating against mock catalogues subject to comparable distance smearing (Section \ref{mock}).
 
\label{final dev of gaia}

To map the observed Gaia $\Delta \phi_{\odot}$ ($8.06^\circ \pm 3.09^{\circ}$) to a deceleration rate, we perform a continuous Bayesian inference across our discrete simulation grid. We first exclude unreliable data points (defined as having an amplitude-to-noise ratio less than $0.5$ or an intrinsic phase error greater than $8.0^\circ$). To interpolate the remaining points of the distance-smeared grid (Figure \ref{fig:heatmap_noisy}), we train two continuous Gaussian Process Regressors (GPR): a primary GP to model the physical $\Delta \phi_{\odot}$ across the $\eta-\Omega_b$ parameter space, and a secondary GP to smoothly model the associated simulation uncertainties.

We formulate a log-likelihood function that directly compares the Gaia observation to the GP-interpolated predictions. This explicitly propagates all sources of uncertainty in quadrature: the observational Gaia error, the local simulation error, and the GP interpolation uncertainty. To constrain the parameter space, we apply physically motivated Gaussian prior for pattern speed based on modern literature estimates: $\Omega_b = 37.0 \pm 6 \text{ km s}^{-1} \text{kpc}^{-1}$ \citep{Portail_2016,2019MNRAS.490.4740B,Sanders_2019, Binney2020, Kawata_2021, Clarke_2022, Zhang2024, 2025MNRAS.541..214D}, for the deceleration rate a uniform prior between $0$ and $0.004$ is used for $\eta$. 

\begin{figure}
    \centering
    \includegraphics[width=\linewidth]{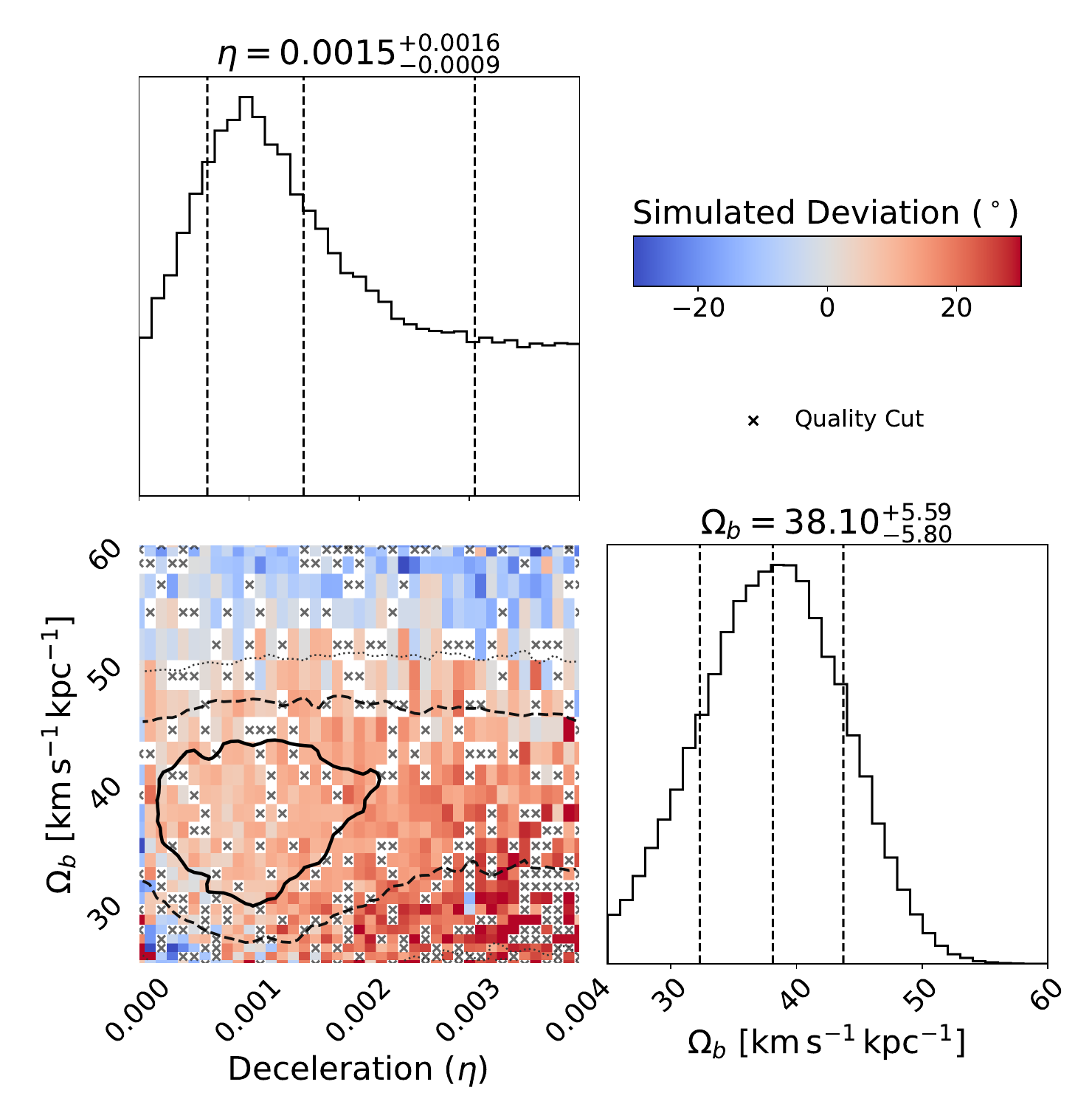}
    \caption{Posterior constraints on $\eta$ and bar pattern speed from comparison of the observed asymmetry phase with the simulation calibration, using the priors specified in Section \ref{observation results}. The lower-left panel shows the joint posterior contours over the discrete simulation $\Delta \phi_{\odot}$ map; crosses mark grid points excluded from GP training by the adopted quality cuts. The upper-left and lower-right panels show the marginal posteriors, with vertical dashed lines marking the 16th, 50th, and 84th percentiles. The marginal estimates are $\eta=0.0015^{+0.0016}_{-0.0009}$ and $\Omega_b=38.10^{+5.59}_{-5.80}\,\mathrm{km\,s^{-1}\,kpc^{-1}}$.}
    \label{fig:final_mcmc_corner}
\end{figure}

We sample the posterior parameter space using an Affine-Invariant Markov Chain Monte Carlo (MCMC) ensemble sampler. The resulting 1D and 2D marginal posterior distributions are presented in Figure \ref{fig:final_mcmc_corner}, overlaid on the distance-smeared simulation grid.

By extracting the 16th, 50th, and 84th percentiles of the converged MCMC chains, we establish kinematic constraints. We constrain the Galactic bar's pattern speed to $\Omega_b = 38.10^{+5.59}_{-5.80} \text{ km s}^{-1} \text{kpc}^{-1}$, and the dimensionless deceleration rate to $\eta \equiv - \frac {\dot{\Omega}_b}{\Omega_b^2} = 0.0015^{+0.0016}_{-0.0009}$. These values are consistent with previous measurements favouring a slow, decelerating Galactic bar \citep{Chiba2021, Zhang_2025}, obtained here from an independent tracer. The marginal posterior on $\eta$ peaks away from zero but retains appreciable probability at $\eta = 0$, and its upper tail is limited by the extent of our simulation grid ($\eta \le 0.004$). The posterior on $\Omega_b$ closely reproduces the adopted prior, so our data do not independently constrain the pattern speed. This is expected: in the idealised simulations $\Delta\phi_\odot$ is governed mainly by $\eta$ (Figure \ref{fig:heatmap}), heliocentric distance errors add a dependence on $\Omega_b$ (Figure \ref{fig:amplitude_heatmap}), and a single measured phase cannot constrain both parameters.

\section{Discussion} \label{discussion}

In galactic dynamics, a rotating bar decelerates primarily by transferring its angular momentum to the dark matter halo and the outer stellar disc via dynamical friction \citep{1984MNRAS.209..729T}. By measuring the deceleration using phase-space chevrons, we are indirectly measuring the friction exerted by the dark matter halo \citep{Debattista_1998}. Because a faster deceleration implies a denser or more centrally concentrated dark matter halo absorbing that momentum \citep{Debattista_2000}, our framework provides a vital diagnostic tool for understanding the underlying mass distribution of the Milky Way.

Our results support the hypothesis that the phase-space chevrons observed in the Milky Way are caused by resonant sweeping from the Galactic bar, rather than being the phase-mixed relics of a past merger \citep{2024MNRAS.532.4389D}. The strong asymmetry observed in the phase-space chevrons in the Gaia DR3 data is difficult to explain with debris from ancient merger events (e.g., $>8 \text{ Gyr}$ ago), as any such kinematic structures would be largely phase-mixed by the present day \citep{1999MNRAS.307..495H, Belokurov_2018, 2024MNRAS.531.1422D}. Instead, the asymmetry pattern is naturally explained by resonant trapping by the bar, and its azimuthal offset is consistent with a decelerating bar.

Using our Bayesian MCMC framework, we extract a dimensionless deceleration parameter of $\eta \equiv - \frac {\dot{\Omega}_b}{\Omega_b^2} = 0.0015^{+0.0016}_{-0.0009}$. This value is smaller than, but in statistical agreement with, the estimate of $\eta = 0.0031 \pm 0.0008$ reported by \citet{Chiba2021}, a value recently verified by \citet{Zhang_2025}. A key distinction between these measurements lies in our larger uncertainties, which we anticipate will be refined with the advent of Gaia DR4.


Equivalently we predict $\dot{\Omega}_b=-2.23^{+1.50}_{-2.46} \, \text{km s}^{-1} \text{kpc}^{-1} \text{Gyr}^{-1}$, whereas \citet{Chiba2021} found $\dot{\Omega}_b = - 4.5 \pm 1.4 \, \text{km s}^{-1} \text{kpc}^{-1} \text{Gyr}^{-1}$. The values in \citet{Chiba2021} were verified by \citet{Zhang_2025}.

Our posterior on the pattern speed, $\Omega_b = 38.10^{+5.59}_{-5.80} \, \text{km s}^{-1} \text{kpc}^{-1}$, is driven by the prior, which is based on previous estimates in the range $32.5$ to $42 \, \text{km s}^{-1} \text{kpc}^{-1}$ \citep{Portail_2016,2019MNRAS.490.4740B,Sanders_2019, Binney2020, Kawata_2021, Clarke_2022, Zhang2024, 2025MNRAS.541..214D}.

\begin{figure}
        \centering
        \includegraphics[width=\linewidth]{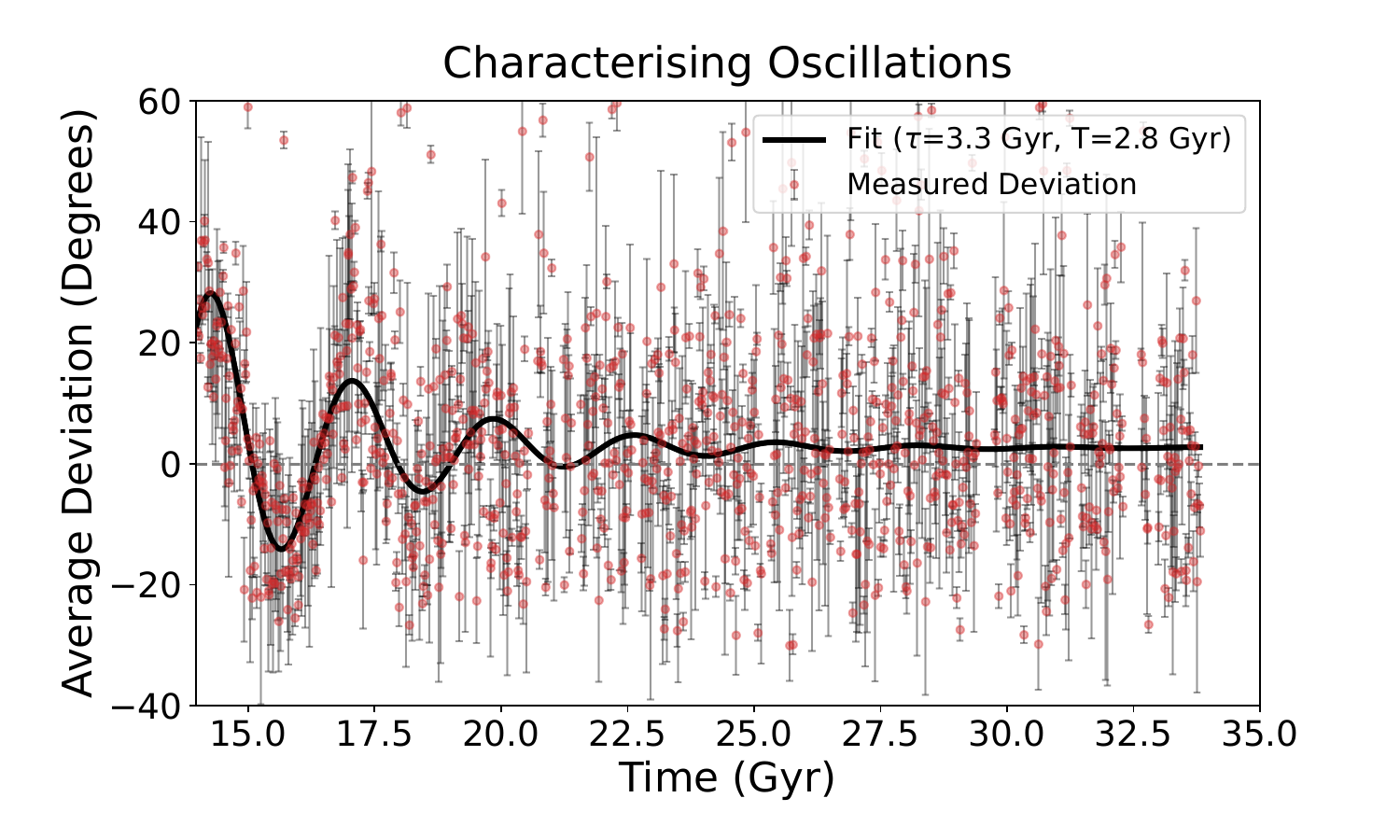}
        \caption{Damped oscillatory response of the measured asymmetry phase after the bar's imposed deceleration ceases. Red points with error bars show the snapshot measurements, and the black curve shows the fitted damped oscillation. The displayed fit gives a period of $T=2.8\,\mathrm{Gyr}$ and a damping time of $\tau=3.3\,\mathrm{Gyr}$. These quantify the oscillation and decay of the response rather than a unique fixed delay between deceleration and measured phase.}
        \label{fig:decel and stop diagnostic2}
\end{figure}

This study assumes $\Delta \phi_{\odot}$ (defined in Section \ref{math_framework}) for the Milky Way is a `live' measurement of the deceleration of the system. In reality the deceleration of the system could be varying \citep{2006ApJ...639..868S}, so a decelerate and stop simulation (as introduced in Section \ref{N-body}) is run to find the lag time between a change in deceleration and the corresponding change in $\Delta \phi_{\odot}$. We run a test simulation (which is presented in Appendix \ref{decelstop}) where the deceleration rate varies as a step function, such that $\eta$ instantly drops from $0.002$ to $0$. Following this drop, we observe an oscillatory response, as shown in Figure \ref{fig:decel and stop diagnostic2}.

The phase oscillates with a period of $T= 2.79 \pm 0.01 \text{ Gyr}$, and the oscillations decay on a time-scale of $\tau = 3.33 \pm 0.06 \text{ Gyr}$. This indicates that the kinematic response retains a memory of the bar's previous deceleration rather than immediately relaxing to its new equilibrium. Consequently, the measured asymmetry may not provide an instantaneous measurement of the bar's current deceleration if its rotational history has varied significantly over the past few billion years.

When we observe the corresponding timestamps the kinematic pattern clearly overshoots the bar axes before settling. The twist therefore encodes the bar's recent deceleration history, not only its present rate: a given $\Delta\phi_\odot$ can be produced by a bar that is decelerating now, or by one whose deceleration has recently changed. Because the response timescale is set by the libration of the trapped orbits, which varies across the resonance, measuring $\Delta\phi$ in several annuli, or at more than one resonance, could in principle separate the present-day deceleration from its recent history. We leave this to future work with Gaia DR4.

While our model successfully reproduces the observed phase-space morphology, we made several simplifying assumptions. Our simulations used a fixed background potential and a mathematically imposed bar deceleration. In reality, the Milky Way has a `live' dark matter halo and a self-gravitating disc that dynamically responds to the bar. Our model also only considers stars; gas dynamics can also affect the evolution and pattern speed of a bar. Further studies could improve upon our simulations by incorporating a more realistic representation of the Galactic system. Our test-particle approach neglects the self-gravity of the halo response; quantifying its effect on $\Delta \phi_{\odot}$ requires live simulations, which we leave to future work.

Two further assumptions affect the quantitative value of $\eta$. First, in the pendulum approximation the equilibrium phase depends on $\eta/A_b$ (equation \ref{shifted eq phase}), so our result is tied to the bar strength of the \citet{Hunter_2024} model; to first order, a bar stronger or weaker by a factor $f$ would imply an $\eta$ larger or smaller by the same factor. Second, our mock catalogue applies a uniform $10\%$ distance error, whereas the median error of the Gaia sample is $6.6\%$; this is physically motivated, but the mock is likely to overestimate the distance smearing, which would bias our inferred $\eta$ high.

Another limitation of this study is the assumption that the bar has a low deceleration rate. However, this is physically reasonable and the presence of a clear kinematic pattern suggests the Milky Way is not one of the high-$\eta$, irregular systems we have ignored. 

We also assume the system does not experience any rapid, dramatic perturbations, for example an interaction with a dwarf galaxy. A merger would also produce phase-space chevrons. However, these structures lack the fractional asymmetry we have probed, so one might naively assume they would escape our analysis technique. Nevertheless, such a dramatic event could also complicate the adiabatic sweeping signature caused by the Galactic bar. 

Future investigations utilising fully self-consistent, live simulations could improve the exploration of these non-linear effects. More immediately, the upcoming release of Gaia DR4 (in December 2026) will dramatically increase the available dataset and significantly reduce line-of-sight distance smearing. Rebuilding our mock catalogues with DR4 constraints will sharpen the azimuthal phase-shift signal, enabling an even more precise measurement of the Milky Way's bar-halo interaction.

\section{Conclusion} \label{conclusion}

The discovery of chevron-shaped phase-space over-densities in Gaia DR3 data has provided a unique window into the dynamics of the Milky Way. Our findings support the hypothesis that these structures are the direct kinematic manifestation of stars trapped in resonances by the Galactic bar. We conducted a set of test particle simulations, parametrising the bar's deceleration via the dimensionless deceleration parameter $\eta$. By adding realistic observational uncertainties to these simulations, we isolated the resonant signatures unique to a decelerating system with the observational artifacts of current data.

Our results demonstrate that a decelerating bar exerts an Euler force on trapped resonant orbits, causing the phase-space asymmetry to twist out of alignment with the Galactic bar's axes. $\Delta \phi_{\odot}$ (defined in Section \ref{math_framework}) scales with the deceleration rate. Applying this framework to the observed Gaia DR3 dataset, we successfully recovered the phase-space chevrons observed in previous studies and extended this analysis to map the chevron asymmetry as a function of position in the Galactic plane. We used this spatial asymmetry to measure $\Delta \phi_{\odot}$ for Gaia DR3 data to be $8.06^\circ \pm 3.09^\circ$. In our mock catalogues (10 per cent distance errors), heliocentric distance errors bias the recovered phase towards the Solar azimuth, most strongly where the intrinsic $m=2$ amplitude is weak (Figure \ref{fig:amplitude_heatmap}); the comparison with Gaia therefore relies on this forward-modelled calibration.

By comparing this observation with our simulated parameter space using a continuous Bayesian MCMC framework, we estimate $\eta \equiv - \frac {\dot{\Omega}_b}{\Omega_b^2} = 0.0015^{+0.0016}_{-0.0009}$. Combined with the pattern-speed posterior (which follows our prior), this corresponds to a current slowing rate of $\dot{\Omega}_b = -2.23^{+1.50}_{-2.46} \, \text{km s}^{-1} \text{kpc}^{-1} \text{Gyr}^{-1}$.

This measurement is consistent with a Galactic bar that is currently decelerating, transferring angular momentum to the outer disc and dark matter halo via dynamical friction. However, our simulations show that the resonant phase-space morphology does not relax immediately when $\eta$ changes: in a test simulation in which $\eta$ drops abruptly from 0.002 to 0, $\Delta\phi_\odot$ overshoots the bar axis and oscillates with a period of $\approx3$ Gyr, damping over $\approx3$ Gyr (Appendix \ref{decelstop}). Consequently, the observed asymmetry and our measurement of $\eta$ could be `out of date' if the deceleration rate of the Milky Way has varied significantly over the past few billions of years.

Ultimately, our methodology establishes chevron spatial asymmetry as a powerful probe of the Milky Way's bar-halo interaction. Future investigations using Gaia DR4 could improve our current estimates.

\section*{Acknowledgements}

AA would like to thank her co-authors for their invaluable guidance, support, and supervision throughout this work. AMD acknowledges support from the Royal Society (URF\textbackslash R1\textbackslash191555; URF\textbackslash R\textbackslash 241030). The authors are grateful to the Institute of Astronomy, University of Cambridge, for providing the high-performance computing resources necessary to execute and analyse the numerical simulations presented in this study.

\section*{Data Availability}

The observational data used in this work is publicly available via the European Space Agency (ESA) Gaia archive (\url{https://gea.esac.esa.int/archive/}). The underlying  simulation data, mock catalogues, and analysis scripts generated during this study are stored on local institutional servers and will be shared upon reasonable request to the corresponding author.



\bibliographystyle{mnras}
\bibliography{bib} 




\appendix

\section{The Decelerate and Stop Case} \label{decelstop}

The deceleration rate of a Galactic bar is not necessarily constant over its lifetime  \citep{2006ApJ...639..868S}. Therefore, we investigate `unphysical' deceleration rates such as $\eta=0.0001$ which would have taken hundreds of billions of years to evolve, as these near-0 deceleration rates could be a recent stage in the galaxy's history.

It is necessary to determine whether the phase-space morphology retains a `memory' of past variations in $\eta$. 

We first ran a test where the deceleration varies gradually. Here the system's kinematics track the instantaneous deceleration rate on the order of billions of years. 

To test the time delay in response, we simulated a scenario where the bar decelerates but eventually stabilizes at a constant rotation speed. The pattern speed varies as:

\begin{equation} \label{decelerate and stop eq}
\Omega_b(t) = 
\begin{cases} 
\Omega_{b,i} & , t_0 \leq t \leq t_1 \\
\Omega_{b,i} \left[ 1 + \frac{1}{2} \eta \Omega_{b,i} (t - t_1)^2 / (t_2 - t_1) \right]^{-1} & , t_1 < t \leq t_2 \\
\Omega_{b,2} [ 1 + \eta \Omega_{b,2} (t - t_2) ]^{-1} & , t_2 < t \leq t_3 \\
\Omega_f & , t > t_3
\end{cases}
\end{equation}

where
\begin{equation*}
\Omega_{b,2} = \Omega_{b,i} \left[ 1 + \frac{1}{2} \eta \Omega_{b,i} (t_2 - t_1) \right]^{-1} ,
\end{equation*}

and
\begin{equation*}
t_3 = t_2 + \frac{1}{\eta} \left( \frac{1}{\Omega_f} - \frac{1}{\Omega_{b,2}} \right) .
\end{equation*}

This simulation begins at constant pattern speed ($\eta=0$), smoothly transitions to a deceleration phase of $\eta=0.002$ between $t_2$ and $t_3$, then drops instantaneously to $\eta=0$ at $t_3$. 

\begin{figure}
        \centering
        \includegraphics[width=\linewidth]{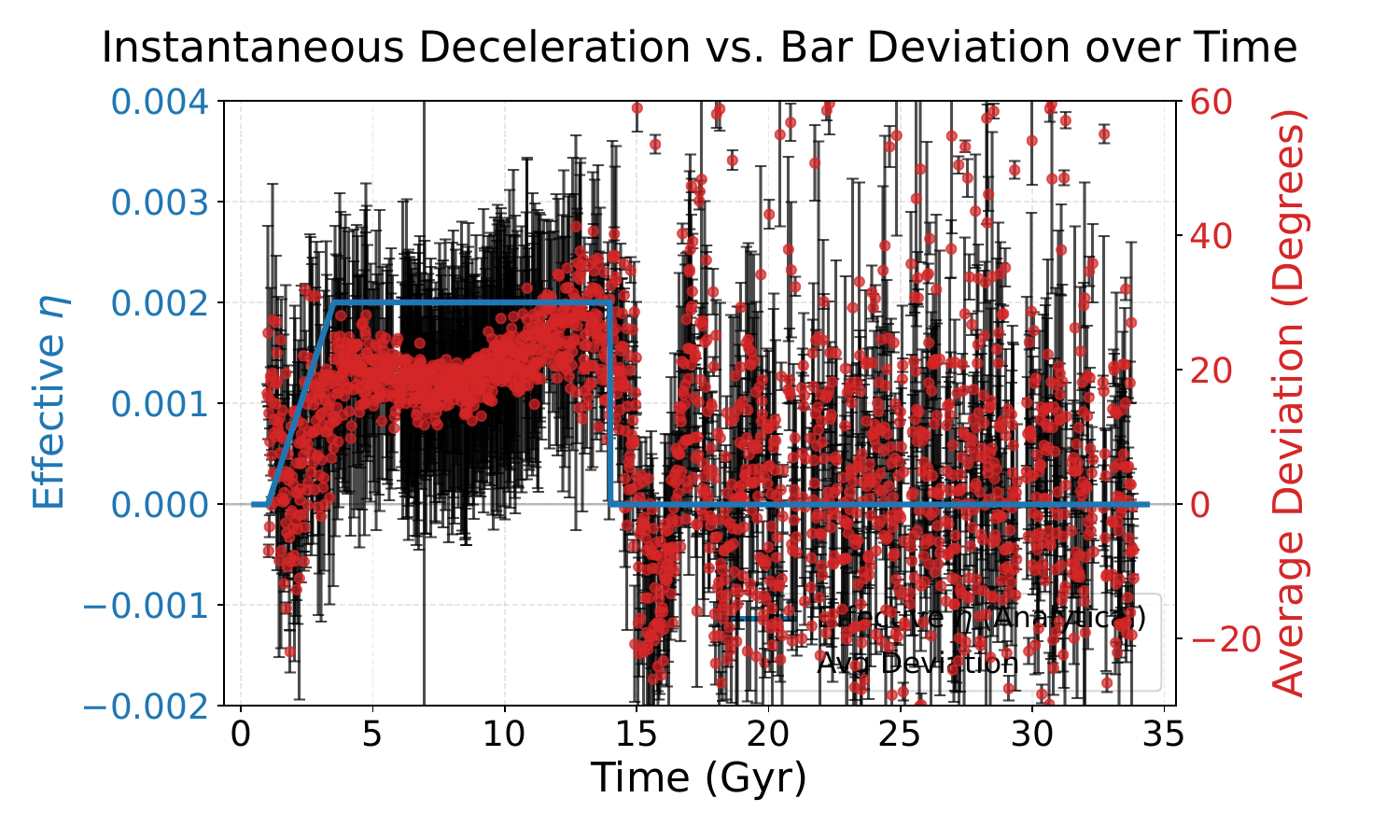}
        \caption{Response of the measured asymmetry phase to a bar that decelerates and subsequently settles to a constant pattern speed. The blue curve shows the imposed instantaneous deceleration parameter on the left axis; red points with error bars show the measured $\Delta \phi_{\odot}$ on the right axis. After the imposed deceleration ends, the phase exhibits an oscillatory response rather than immediately returning to the bar-axis alignment. Simulation time is measured from the adopted initialisation, as defined in Appendix \ref{decelstop}.}
        \label{fig:decel and stop diagnostic}
\end{figure}

Figure \ref{fig:decel and stop diagnostic} quantifies how quickly $\Delta \phi_{\odot}$ (defined in Section \ref{math_framework}) responds to this instantaneous halt. 

What we see here is $\Delta \phi_{\odot}$ varying as expected until the sudden change, after this it appears to start oscillating and then enters into an erratic regime. To analyse this we used an MCMC to fit an oscillating signal to the data after $t_3$; this produces Figure \ref{fig:decel and stop diagnostic2}.

\section{The Accelerating Case} \label{accel}

So far we have considered $\eta \geq 0$. For completeness, we also examined the phase-space response to an accelerating bar ($\dot{\Omega}_b > 0$ and $\eta<0$). We executed a simulation mirroring the setup of our decelerating models, but with a constant acceleration rate of $\eta = -0.002$.

Mathematically, an accelerating bar inverts the sign of the driving term in the pendulum approximation (equation \ref{shifted eq phase}). The dimensionless deceleration parameter $\eta$ becomes negative, shifting the equilibrium phase $\theta_{\mathrm{eq}}$ in the opposite direction compared to the decelerating case. Physically, an accelerating bar causes the resonance to sweep inward, shrinking the phase-space trapping volume. Rather than trapping and dragging stars to create over-densities, the shrinking resonance expels stars, returning them to untrapped orbits and leaving phase-space under-densities.

\begin{figure}
    \centering
    
    \begin{subfigure}[b]{\columnwidth}
        \centering
        \includegraphics[width=\linewidth]{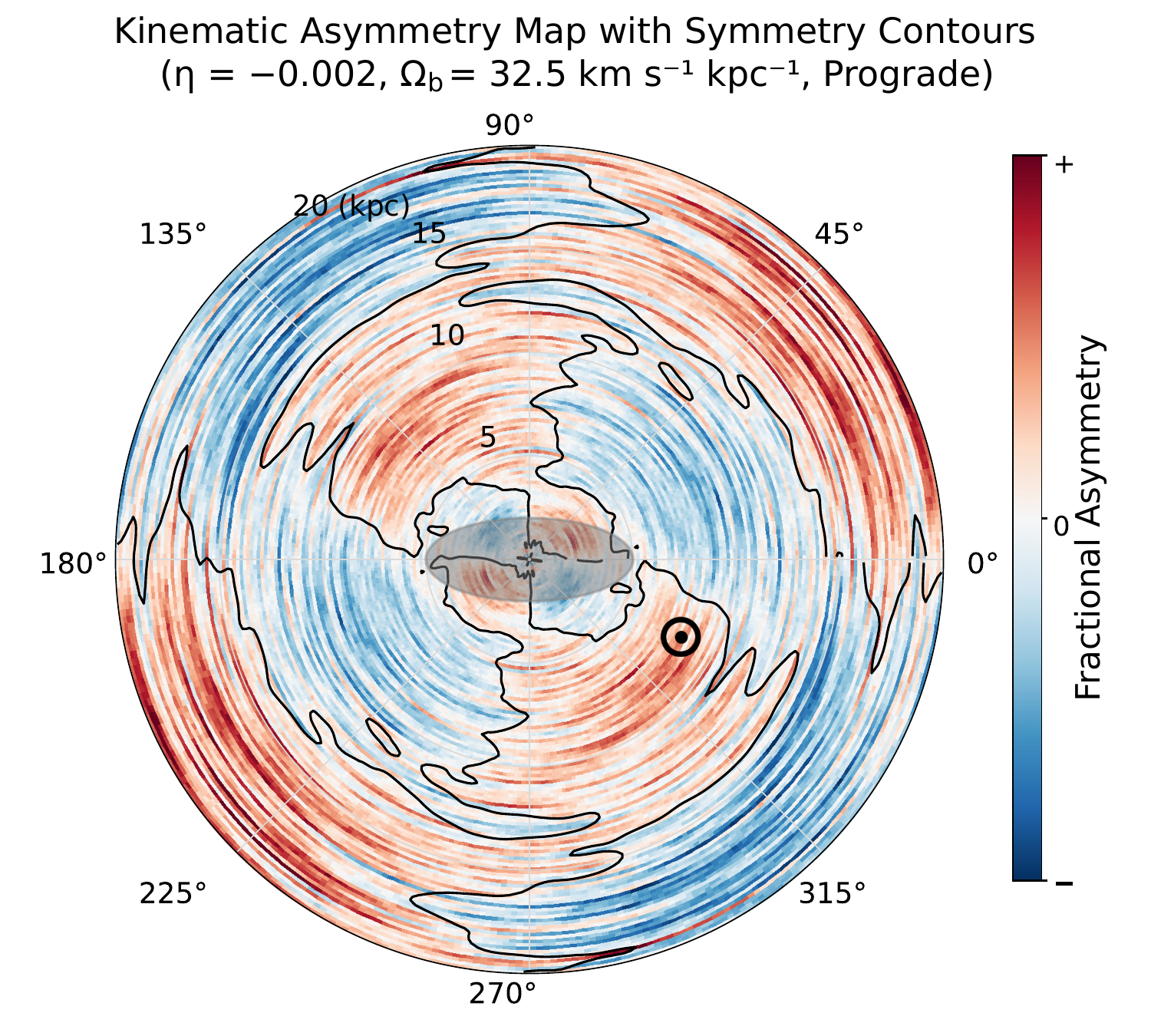}
        \caption{Spatial asymmetry signal for the accelerating system at $\Omega_b=32.5\,\mathrm{km\,s^{-1}\,kpc^{-1}}$, before streaming dominates.}
        \label{fig:accel_1}
    \end{subfigure}

    \begin{subfigure}[b]{\columnwidth}
        \centering
        \includegraphics[width=\linewidth]{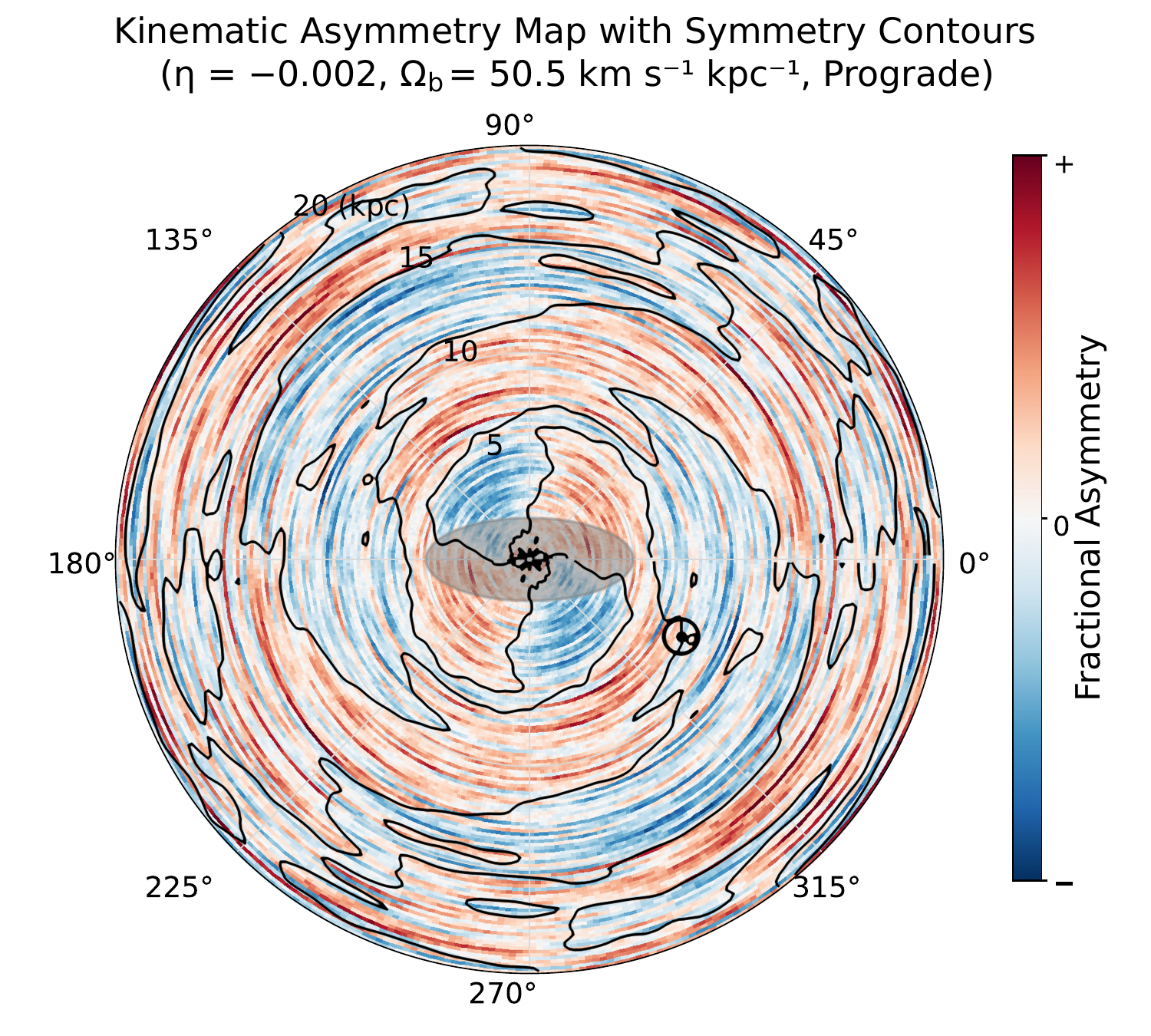}
        \caption{Spatial asymmetry signal for the accelerating system at $\Omega_b=50.5\,\mathrm{km\,s^{-1}\,kpc^{-1}}$, after streaming dominates.}
        \label{fig:accel_2}
    \end{subfigure}
    
    \caption{Spatial radial-velocity asymmetry in the prograde population of the accelerating-bar simulation, with $\eta=-0.002$. Panels (a) and (b) show snapshots at $\Omega_b=32.5$ and $50.5\,\mathrm{km\,s^{-1}\,kpc^{-1}}$, respectively, as indicated in the plot titles. Red and blue indicate outward and inward excesses; black curves show contours derived from the smoothed maps, and the grey ellipse marks the bar. The orientation and morphology differ from those of the decelerating simulation shown in Figure \ref{fig:smoothed_contours_combined}.}

    \label{fig:accel_sequence_combined}
\end{figure}

Figure \ref{fig:accel_sequence_combined} displays two examples of spatial asymmetry snapshots at different times in the spin-up galaxy's lifetime.

As expected, the asymmetry signature curves in the opposite direction compared to the decelerating case. Additionally, Figure \ref{fig:accel_1} appears to have the same magnitude of $\Delta \phi_{\odot}$ (defined in Section \ref{math_framework}) as for the decelerating case with $\eta=0.002$. Despite having the same absolute magnitude of $\eta$, the signature in Figure \ref{fig:accel_2} does not exhibit the same magnitude of $\Delta \phi_{\odot}$ as Figure \ref{fig:smoothed_contours_combined}; the amount of deviation from the bar axes is significantly larger and is closer to a spiral. The expulsion of stars from resonance causes streaming of stars, meaning the signature we observe is driven by streaming rather than strictly being an effect of the Euler force. 

\begin{figure}
    \centering
    
    \begin{subfigure}[b]{\columnwidth}
        \centering
        \includegraphics[width=\linewidth]{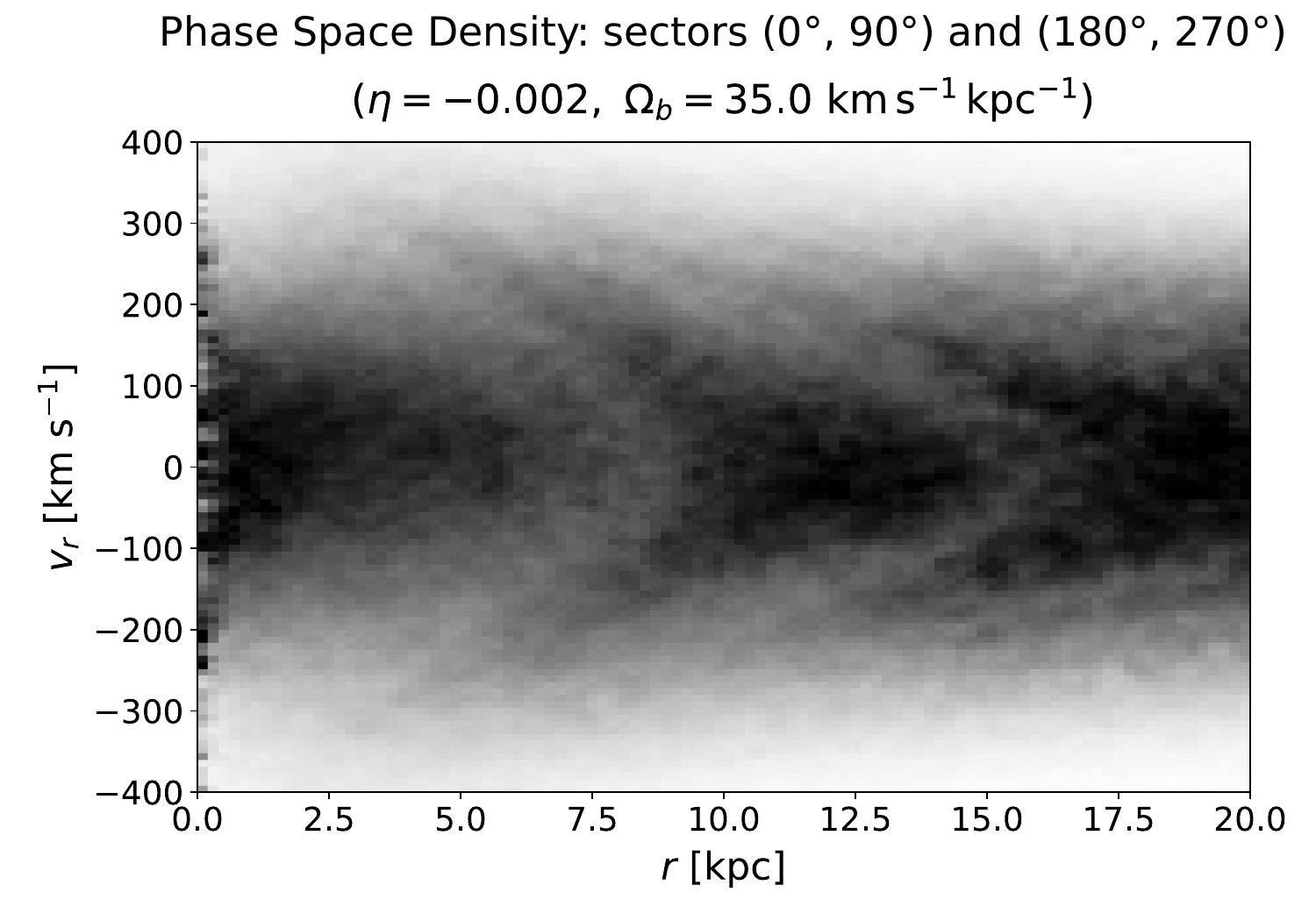}
        \caption{Accelerating bar ($\eta=-0.002$) with $\Omega_b = 35.0 \, \text{km s}^{-1} \text{kpc}^{-1}$.}
        \label{fig:accel_anti_chevron}
    \end{subfigure}

    \begin{subfigure}[b]{\columnwidth}
        \centering
        \includegraphics[width=\linewidth]{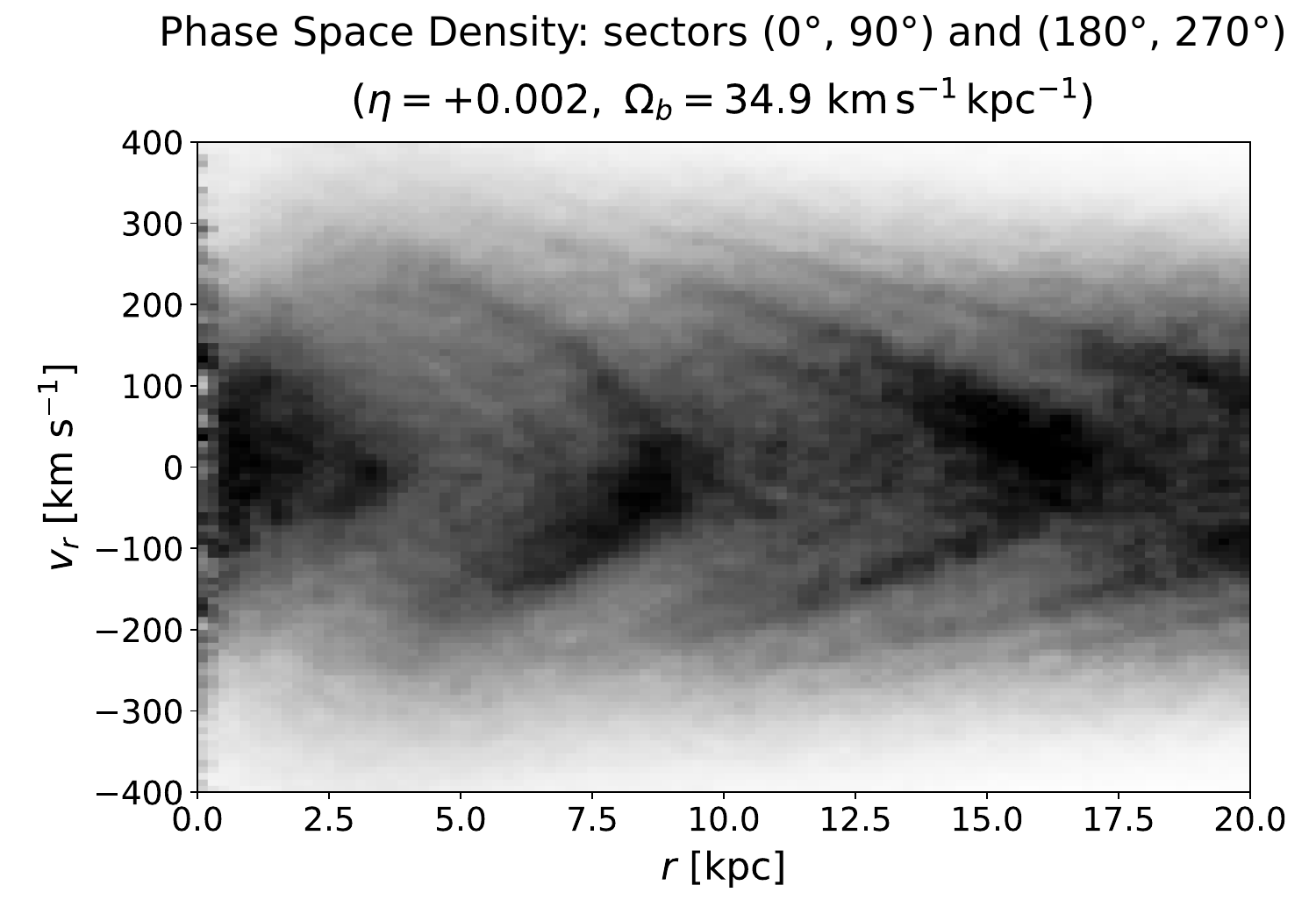}
        \caption{Decelerating bar ($\eta=0.002$) with $\Omega_b = 34.9 \, \text{km s}^{-1} \text{kpc}^{-1}$.}
        \label{fig:decel_comparison_chevron}
    \end{subfigure}
    
    \caption{Comparison of radial phase-space structure in accelerating and decelerating bar models with equal absolute deceleration parameter. The upper panel shows $\eta=-0.002$ at $\Omega_b=35.0\,\mathrm{km\,s^{-1}\,kpc^{-1}}$; the lower panel shows $\eta=+0.002$ at $\Omega_b=34.9\,\mathrm{km\,s^{-1}\,kpc^{-1}}$. Both combine the azimuthal sectors $(0^\circ, 90^\circ)$ and $(180^\circ, 270^\circ)$. The differing distributions illustrate the dependence of the chevron morphology on the imposed bar-evolution history.}
    \label{fig:accel_vs_decel_comparison}
\end{figure}

In phase-space, this expulsion of stars produces distinct `anti-chevrons' characterised by localised phase-space under-densities. A direct comparison between an accelerating and decelerating system is presented in Figure \ref{fig:accel_vs_decel_comparison} where both snapshots have the same magnitude of $\eta$ and similar speeds.

The distinct morphology of these anti-chevrons provides a useful diagnostic. In the Milky Way's observational data, we seem to observe standard over-density chevrons, this can be implied from the width and shape of the over-dense regions compared to the under-dense regions. 

This disfavours scenarios in which the Galactic bar is currently spinning up (e.g., via significant angular momentum gain from gas inflows), and is consistent with secular deceleration driven by dynamical friction with the dark matter halo \citep{1984MNRAS.209..729T}. The positive sign of the measured $\Delta\phi_\odot$ is also consistent with this: it matches the sense of twist in our decelerating models, is opposite to that seen in our accelerating model (Figure \ref{fig:accel_1}), and is opposite to the shift introduced by distance errors.

\section{The Gaia-like Signature} \label{gaia fitting}

After applying the distance smear, we applied a spatial selection function to mimic the precise observational footprint of our Gaia dataset. Using a k-d tree algorithm, we matched the smeared simulated particles to the exact 3D spatial coordinates ($x, y, z$) of our real Gaia targets via a nearest-neighbour query. By discarding any simulated particles that did not correspond to a real observation point, we generated a mock phase-space catalogue. This catalogue intrinsically inherits the complex spatial boundaries, selection biases, and density distribution of the true Gaia DR3 dataset. In principle, this allows a like-for-like comparison with the observed sample.

Subsequently, to mimic the spatial distribution of Gaia data and selection biases of our observational sample, we applied the exact Gaia DR3 spatial selection function (as outlined in Section \ref{mock catalogue}). 

\begin{figure}
        \centering
        \includegraphics[width=\linewidth]{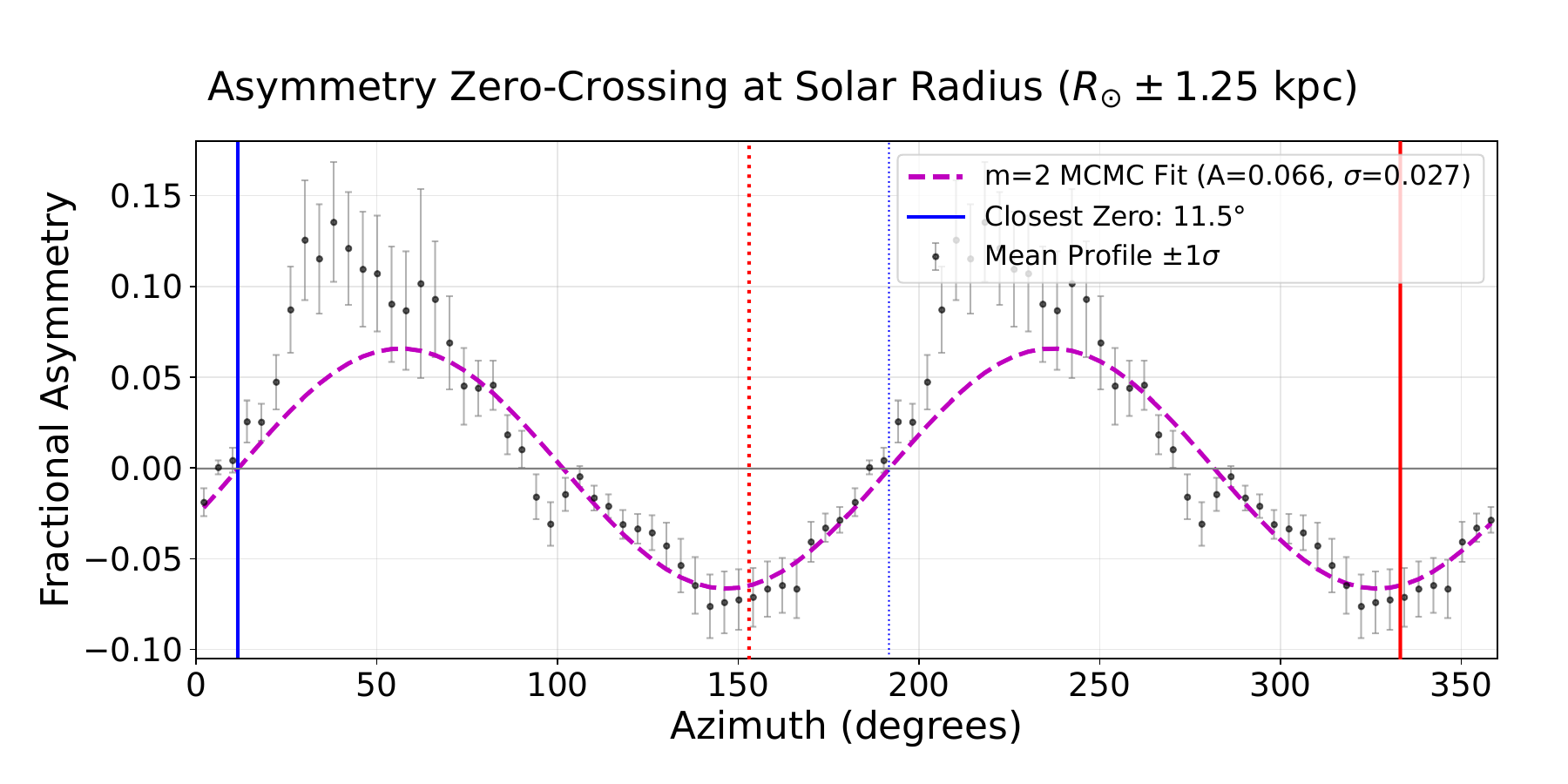}
        \caption{Azimuthal radial-velocity asymmetry in the Solar annulus for a simulated catalogue with both heliocentric distance errors and the Gaia-like spatial matching described in Appendix \ref{gaia fitting}. Points with error bars show the measured profile using the coarser azimuthal binning adopted for this catalogue; the magenta dashed curve is the fitted $m=2$ sinusoid. Blue lines mark the selected zero crossings, and red lines mark the Solar azimuth and its folded counterpart. The displayed crossing is 12.8 degrees.}
        \label{fig:gaia asym curve}
\end{figure}

After taking a Gaia-like distribution cut to the noisy plots, we use the same process to fit the asymmetry curve, but we reduce the number of azimuthal bins. This reduces the resolution of the sinusoidal curves but helps the fitting process. An example is in Figure \ref{fig:gaia asym curve}.

This is the same data as in Figure \ref{fig:dev noisy} but with many of the data points removed; most of the data is lost at the point furthest from the Sun (where the distance smearing has the largest effect) so the largest changes from the previous curve are around $\phi = 63^{\circ}$ and $\phi = 243^{\circ}$.

\begin{figure}
        \centering
        \includegraphics[width=\linewidth]{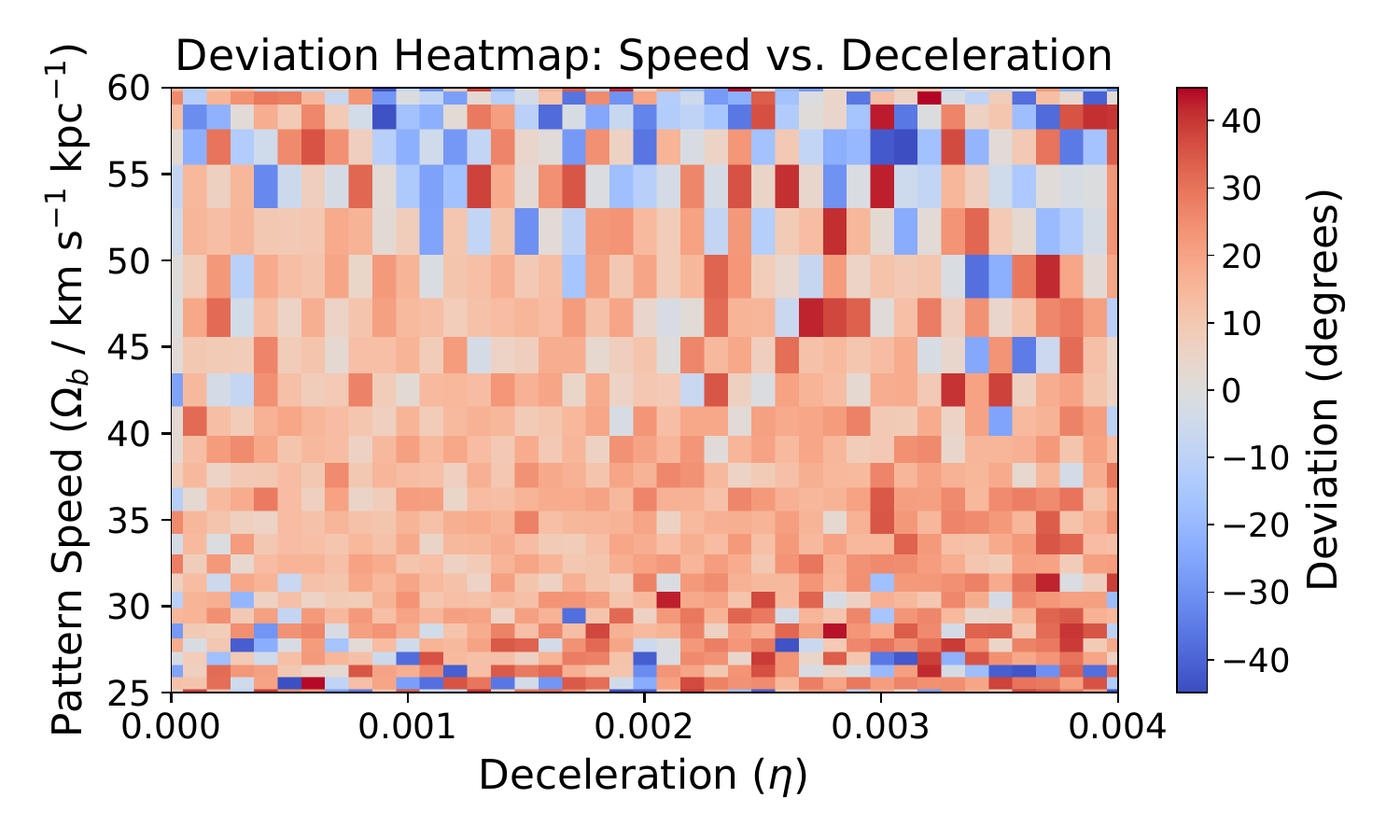}
        \caption{Recovered $\Delta \phi_{\odot}$ in the Solar annulus across the simulation grid after distance smearing and Gaia-like spatial matching. Colours indicate $\Delta \phi_{\odot}$ in degrees. The additional spatial selection produces a less regular calibration than the distance-smeared grid in Figure \ref{fig:heatmap_noisy}; the resulting loss of usable information is discussed in Appendix \ref{gaia fitting}.}
        \label{fig:heatmap_gaia}
\end{figure}

The same process is done to all snapshots across deceleration-speed space; this produces the $\Delta \phi_{\odot}$ (defined in Section \ref{math_framework}) heatmap in Figure \ref{fig:heatmap_gaia}.

With the Gaia selection applied, the calibration in Figure \ref{fig:heatmap_gaia} becomes considerably noisier, because the matched mock retains at most one particle per Gaia star. We therefore adopt the distance-smeared model (Figure \ref{fig:heatmap_noisy}) as our baseline, noting that the Gaia footprint may introduce an additional systematic in $\Delta \phi_{\odot}$ that we do not model.

\section{The Effect of Resonance Swapping on Measuring Deviation (Addressing Heatmap Noise)} \label{transition zones}

\begin{figure} 
        \centering
        \includegraphics[width=\linewidth]{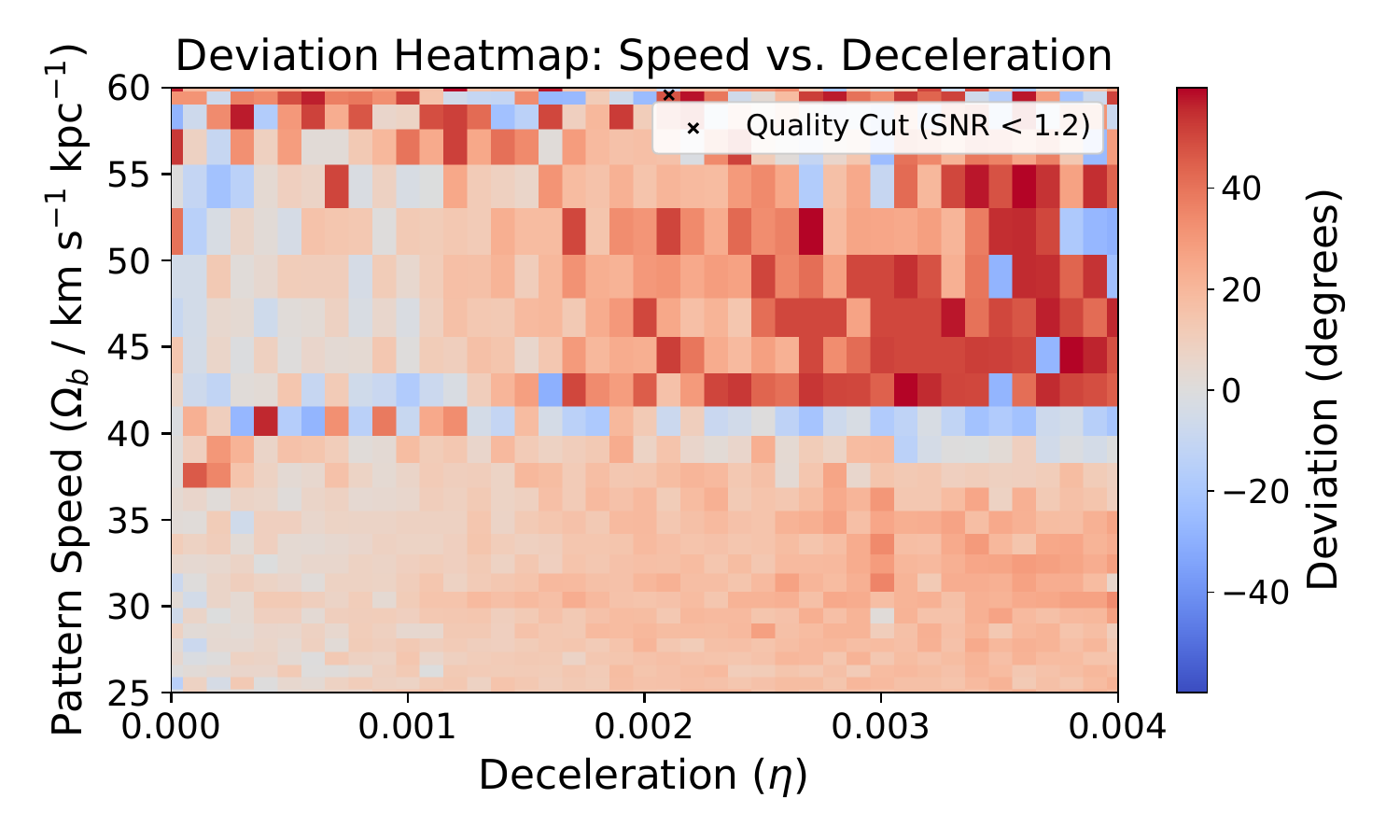}
        \caption{Recovered $\Delta \phi_{\odot}$ measured in an annulus centred at $1.5R_\odot$, across the $\eta$--pattern-speed grid. Colours indicate $\Delta \phi_{\odot}$ in degrees and crosses mark fits below the displayed signal-to-noise threshold of 1.2. The change in phase near pattern speeds of $40$--$43\,\mathrm{km\,s^{-1}\,kpc^{-1}}$ is associated with a change in the asymmetry boundaries sampled by this outer annulus, illustrated in Figure \ref{fig:transition_zone_crossing_combined}.}
        \label{fig:heatmap_1p5R} 
\end{figure}

Returning to the masked data points present in our raw simulation heatmap in Figure \ref{fig:heatmap}, we investigated how transition zones between adjacent resonances affect the extraction of $\Delta \phi_{\odot}$ (defined in Section \ref{math_framework}). To probe this effect, we repeated our analysis but shifting the central radius of the annulus to $1.5 \times R_{\odot}$ from $R_{\odot}$. The resulting $\Delta \phi_{\odot}$ heatmap is presented in Figure \ref{fig:heatmap_1p5R}.

\begin{figure}
    \centering
    
    \begin{subfigure}[b]{\columnwidth}
        \centering
        \includegraphics[width=\linewidth]{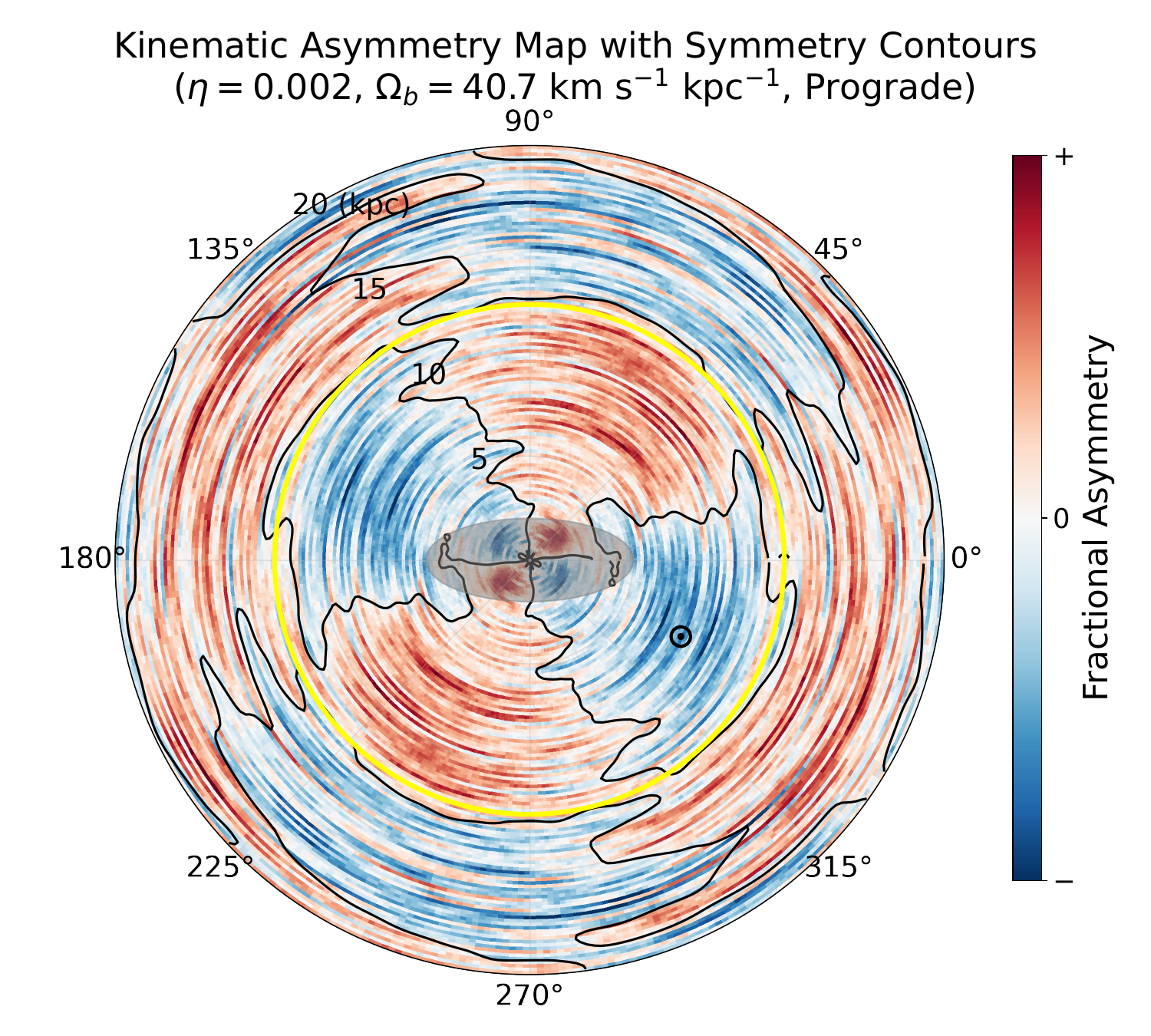}
        \caption{$\Omega_b = 40.7 \, \text{km s}^{-1} \text{kpc}^{-1}$ (before crossing)}
        \label{fig:polar_smooth_40}
    \end{subfigure}
 
    \begin{subfigure}[b]{\columnwidth}
        \centering
        \includegraphics[width=\linewidth]{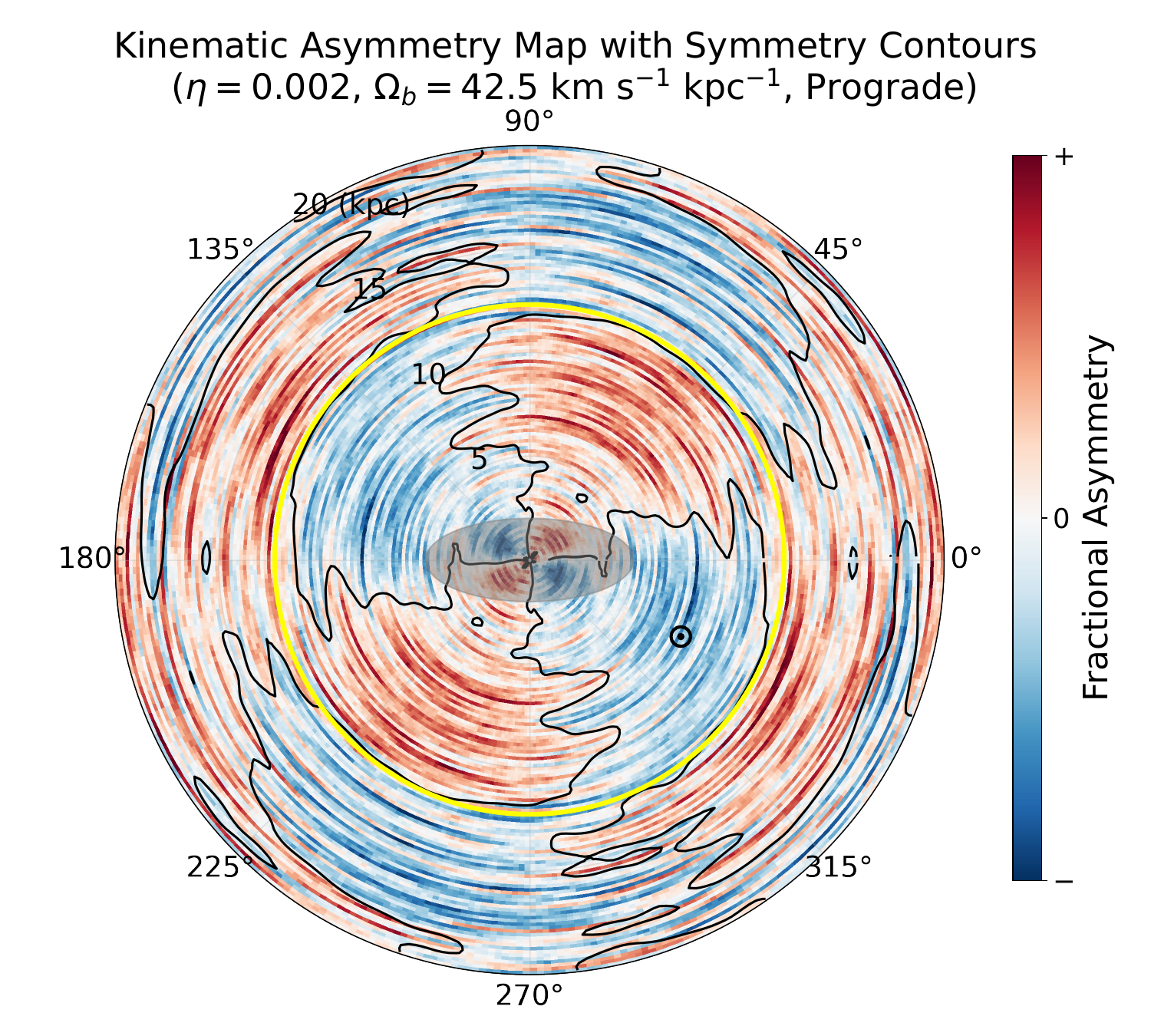}
        \caption{$\Omega_b = 42.5 \, \text{km s}^{-1} \text{kpc}^{-1}$ (after crossing)}
        \label{fig:polar_smooth_42} 
    \end{subfigure}
    
    \caption{Spatial radial-velocity asymmetry at $\eta=0.002$ for snapshots with (a) $\Omega_b=40.7$ and (b) $42.5\,\mathrm{km\,s^{-1}\,kpc^{-1}}$. The highlighted annulus is centred at $1.5R_\odot$. Changes in the locations of the smoothed asymmetry boundaries relative to this annulus help explain the phase transition in Figure \ref{fig:heatmap_1p5R}. Red and blue indicate outward and inward excesses; the grey ellipse marks the bar.}
    \label{fig:transition_zone_crossing_combined}
\end{figure}

This map has a discontinuity between $\Omega_b = 40.75 \, \text{km s}^{-1} \text{kpc}^{-1}$ and $\Omega_b = 42.54 \, \text{km s}^{-1} \text{kpc}^{-1}$. Observing the corresponding spatial asymmetry maps (Figure \ref{fig:transition_zone_crossing_combined}) shows that the physical boundary between two distinct resonance structures crosses the targeted $1.5 R_\odot$ ($12.3 \text{ kpc}$) annulus exactly at these speeds.

Because the extraction annulus simultaneously encapsulates both resonances in this transition zone, the macroscopic $m=2$ sinusoidal signature temporarily breaks down. This localised interference prevents a clean extraction of $\Delta \phi_{\odot}$, creating the unconstrainable data points observed across our heatmaps. Crucially, the most prominent transition zones do not cross the Solar radius within our sampled range of pattern speeds. Although a fainter resonance boundary does intersect $R_{\odot}$ driving the specific measurement difficulties observed at $\eta=0$ and $\Omega_b \approx 32.5 \text{ km s}^{-1} \text{kpc}^{-1}$, the severe discontinuity observed in the $1.5 R_{\odot}$ heatmap avoids our primary extraction annulus.

To mitigate the impact of this type of non-sinusoidal noise on our primary results, we naturally accommodated minor $\Delta \phi_{\odot}$ using the intrinsic scatter parameter ($\sigma$) in our MCMC log-likelihood. Also we mask any data points exhibiting severe phase errors prior to our final Bayesian interpolation.

\section{The Effect of Bar Length on Results} \label{appendix_bar_length}

The length of the Milky Way's bar remains uncertain \citep{Hilmi2020}, with literature estimates ranging from $L_b\approx 5\,\text{kpc}$ \citep{Wegg_2015, Shen_2020} to $L_b\approx 3.5\,\text{kpc}$ \citep{Lucey_2023}. A strict upper limit of $\approx 8\,\text{kpc}$ is naturally imposed by the solar radius.

In our simulations, the bar length varies according to the pattern speed to reflect realistic evolution; specifically, the bar lengthens as the pattern speed decreases. This is calibrated so that the bar reaches $L_b\approx 5\,\text{kpc}$ at $\Omega_b \approx 35 \, \text{km s}^{-1} \text{kpc}^{-1}$, aligning with standard literature estimates of the current state of the Milky Way bar. Consequently, the bar length varies between snapshots. Given the large observational uncertainties associated with the dimensions of the bar, we must take these errors in modelled and predicted bar lengths into our error analysis.

\begin{figure}
    \centering
        \includegraphics[width=\linewidth]{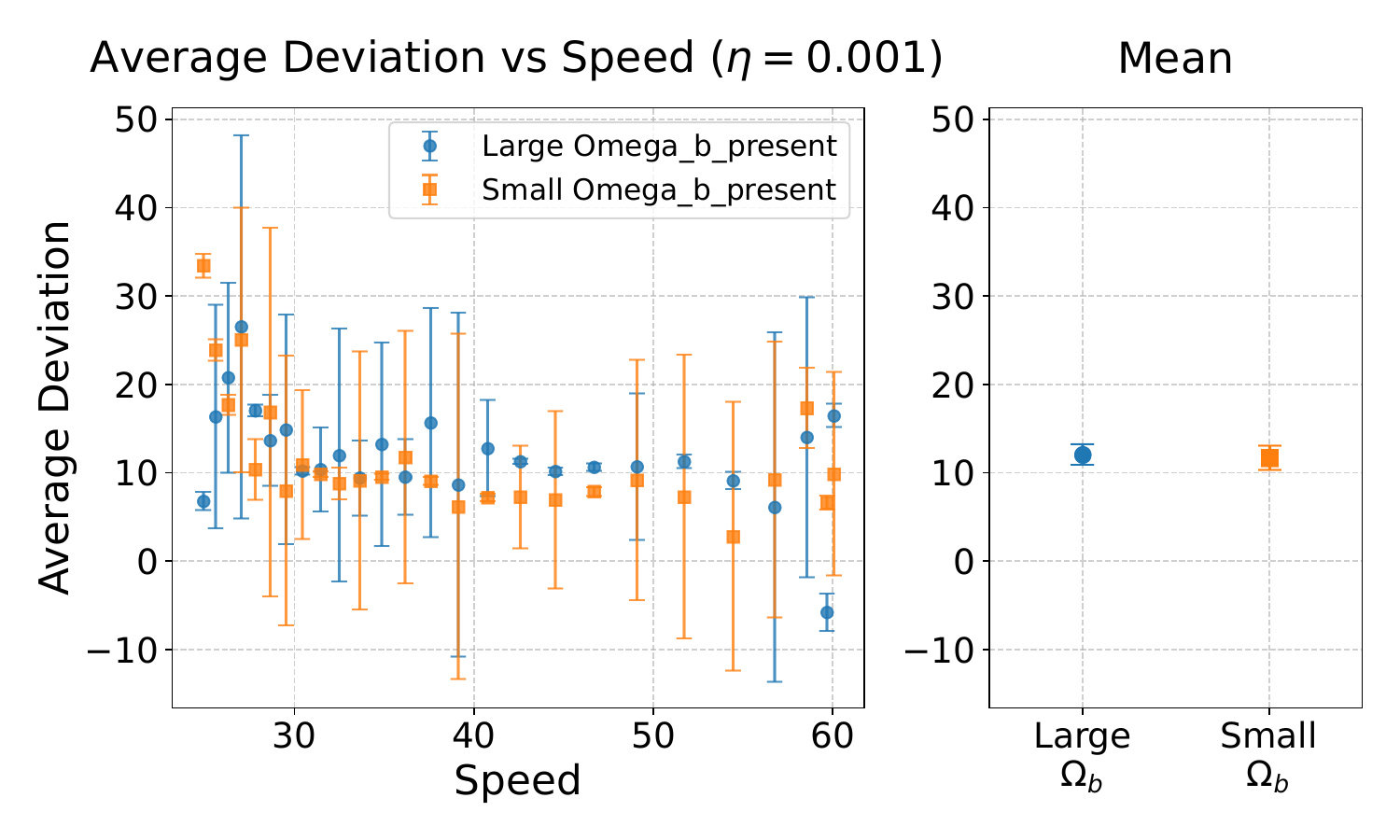}
        \caption{Comparison of measured $\Delta \phi_{\odot}$ for two simulation normalisations labelled by their adopted present-day bar pattern speed, at $\eta=0.001$. The left panel shows $\Delta \phi_{\odot}$ as a function of instantaneous pattern speed for the two cases, with measurement uncertainties; the right panel compares the corresponding mean $\Delta \phi_{\odot}$. The two mean estimates are similar within the displayed uncertainties.}
        \label{fig:bar length test}
\end{figure}

Variations in the bar length only act as a second order effect on $\Delta \phi_{\odot}$ (defined in Section \ref{math_framework}). To quantify this uncertainty, we conducted a test by varying the bar length for the $\eta=0.001$ simulation and compared $\Delta \phi_{\odot}$ before and after for each snapshot across the lifetime of the simulation. This test produced the results in Figure \ref{fig:bar length test}.

These results demonstrate that altering the bar length causes a marginal change in $\Delta \phi_{\odot}$ of the resonant orbit structure, with the measured $\Delta \phi_{\odot}$ remaining within $1 \sigma$ of each other. This introduces a mean shift in $\Delta \phi_{\odot}$ of $1.45^\circ$ at $\eta = 0.001$, computed as the weighted mean of the per-snapshot differences. There does not appear to be a clear trend in speed, and we would not predict one. The speed dependent noise will be encapsulated in the error already associated with high and low speed snapshots. For this $\Delta \phi_{\odot}$ we use an additional error to be added in quadrature to the errors in all the $\eta=0.001$ snapshots.

\begin{figure}
    \centering
        \includegraphics[width=\linewidth]{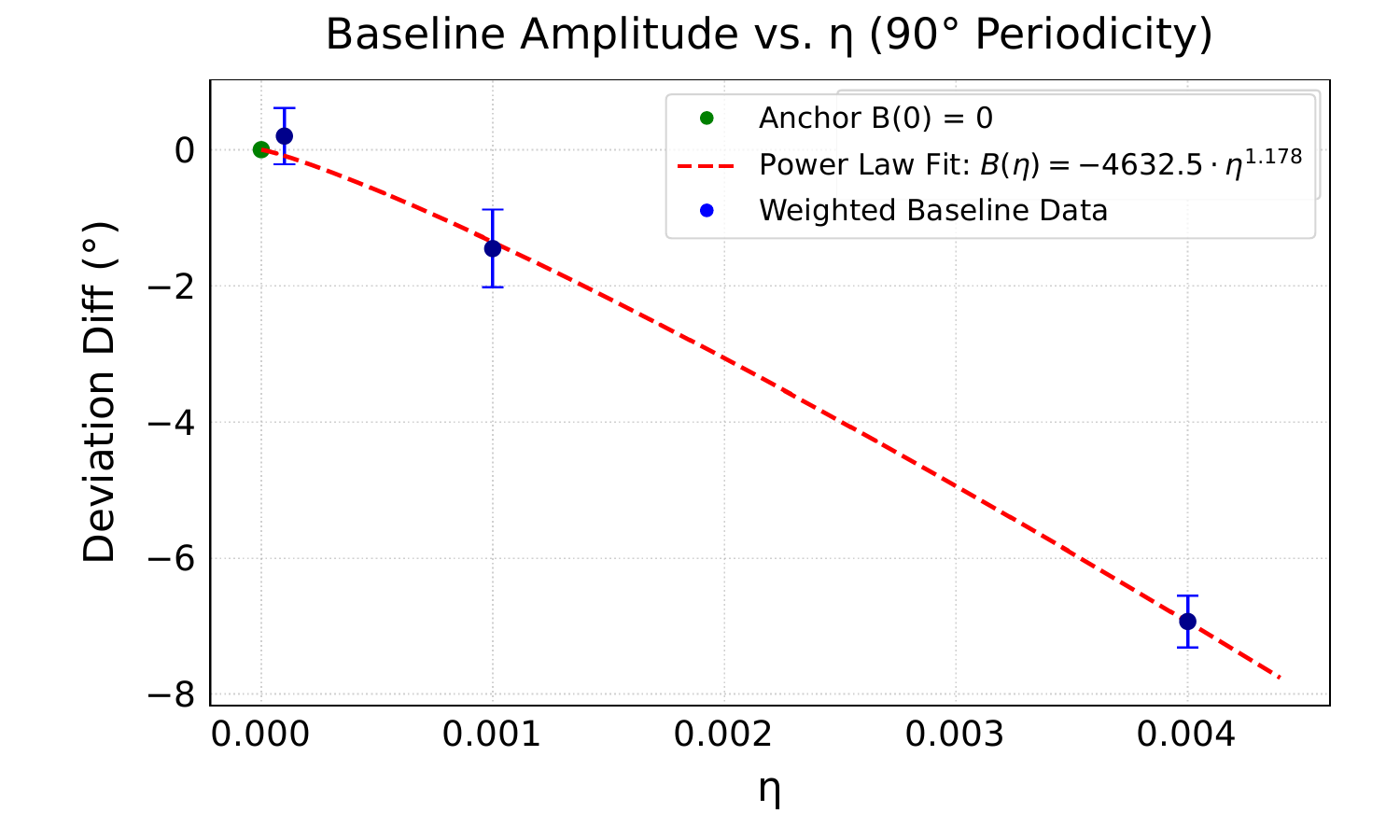}
        \caption{Weighted mean difference in recovered $\Delta \phi_{\odot}$ between the two simulation normalisations compared in Figure \ref{fig:bar length test}, plotted against $\eta$. Blue points with error bars show the weighted baseline measurements, the green point marks the imposed zero-deceleration anchor, and the red dashed curve shows the fitted relation $B(\eta)=-4632.5\,\eta^{1.178}$, with B in degrees. The phase differences are evaluated using the 90-degree periodic convention indicated in the plot title.}
        \label{fig:bar length test2}
\end{figure}

To see how this varies with $\eta$ we run two more simulations with longer bars, one for $\eta=0.0001$ and one for $\eta=0.004$. Since $\Delta \phi_{\odot}=0$ when $\eta = 0$ this will not be changed by a longer or shorter bar, so we use this as an anchor point. We also expect the amount of effect the bar length has increases with $\Delta \phi_{\odot}$. Plotting the difference in $\Delta \phi_{\odot}$ against $\eta$ yields Figure \ref{fig:bar length test2}.

We can interpolate to find an error associated with uncertainty in bar length for each $\eta$ simulation. 

Because these variations are minimal, we incorporate it into our global error budget across all data stages. Changing the bar-length normalisation shifts $\Delta \phi_{\odot}$ systematically downwards, by $\approx 1.5^\circ$ at $\eta = 0.001$ rising to $\approx 7^\circ$ at $\eta = 0.004$; we include this shift in our systematic error budget.

\section{Errors Associated with Data Points} \label{strength}

The amount of noise and the strength of the signal can be represented in a few ways. Firstly we show the amplitude of the fractional asymmetry fit in Figure \ref{fig:amplitude_heatmap}.

The success of the MCMC fit is strongly correlated with the intrinsic strength of the resonant signature. Figure \ref{fig:amplitude_heatmap} maps the fitted $m=2$ amplitude across the parameter space. The amplitude of the signal peaks near $\eta=0.001$. At extremely low deceleration rates, the resonance sweeps too slowly to capture a significant stellar population, whereas at higher rates the sweeping occurs too rapidly for stars to remain adiabatically trapped. 

The amplitude also peaks near $\Omega_b = 35 \, \text{km s}^{-1} \text{kpc}^{-1}$. At high speeds, the bar has not had sufficient evolutionary time to sweep up halo stars. Conversely, at low speeds the resonances migrate past the Solar radius, shifting the macroscopic chevron features entirely out of the observational annulus.

\begin{figure}
    \centering
    \begin{subfigure}[b]{\columnwidth}
        \centering
        \includegraphics[width=\linewidth]{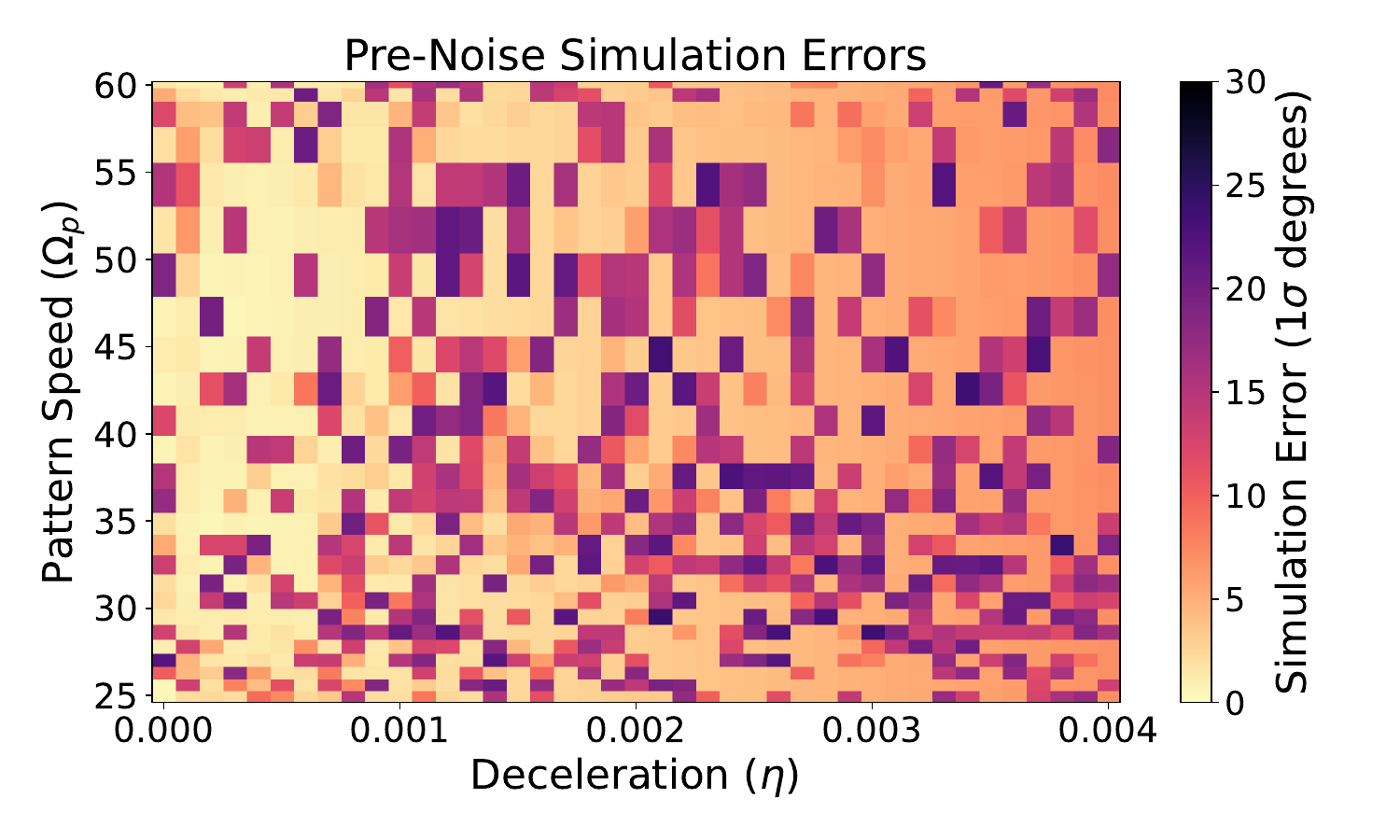}
        \caption{Errors associated with each snapshot's $\Delta \phi_{\odot}$ measurement for the raw simulated data.}
        \label{fig:errors}
    \end{subfigure}
    
    \begin{subfigure}[b]{\columnwidth}
        \centering
        \includegraphics[width=\linewidth]{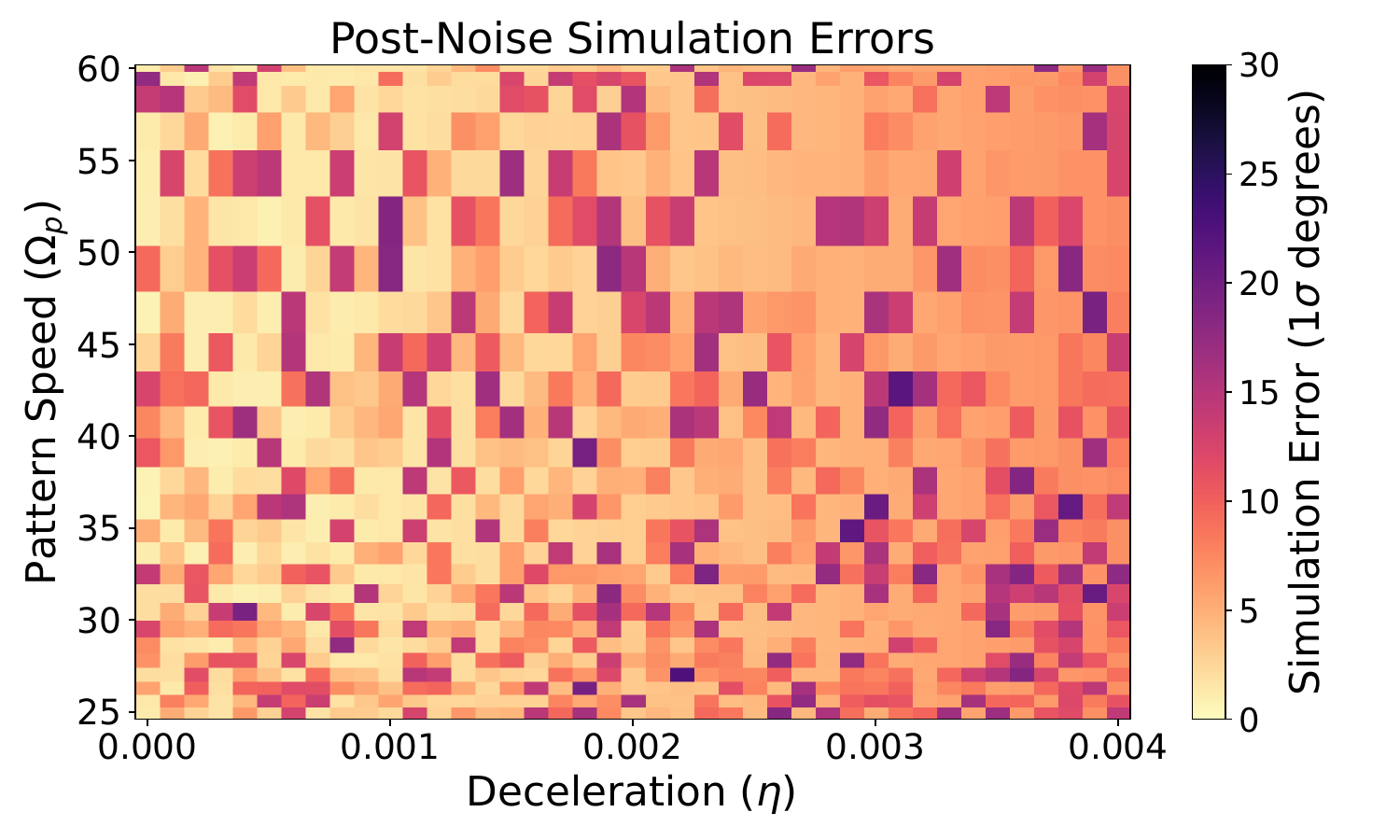}
        \caption{Errors associated with each snapshot's $\Delta \phi_{\odot}$ measurement for the distance-smeared simulated data.}
        \label{fig:errors noisy}
    \end{subfigure}
 
    \begin{subfigure}[b]{\columnwidth}
        \centering
        \includegraphics[width=\linewidth]{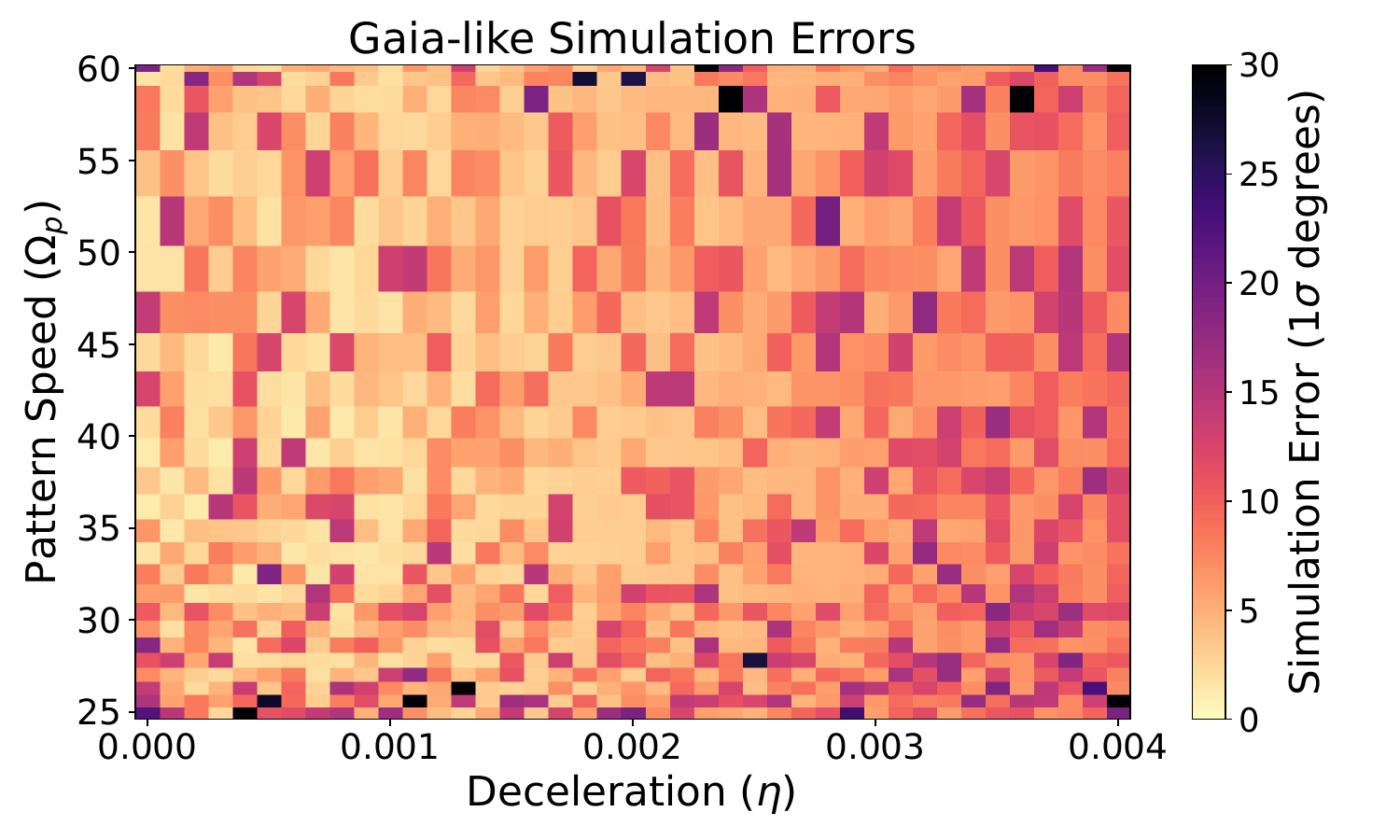}
        \caption{Errors associated with each snapshot's $\Delta \phi_{\odot}$ measurement for the distance-smeared simulated data after a Gaia-like distribution cut.}
        \label{fig:errors gaia}
    \end{subfigure}
    
    \caption{Posterior uncertainty in the recovered asymmetry phase across the simulation grid at three catalogue stages: (a) idealised simulations before distance errors; (b) simulations after heliocentric distance errors; and (c) distance-smeared simulations with additional Gaia-like spatial matching. The common colour scale gives the reported $1\sigma$ phase uncertainty in degrees. These maps describe the statistical precision of the fitted phase, rather than the signed phase bias induced by observational modelling.}
    \label{fig:error dist}
\end{figure}

The statistical uncertainty associated with our measured $\Delta \phi_{\odot}$ (defined in Section \ref{math_framework}) is intrinsically linked to both the physical amplitude of the phase-space asymmetry and its vulnerability to line-of-sight distance smearing (Figure \ref{fig:amplitude_heatmap}). To fully contextualise the reliability of our simulation grid, Figure \ref{fig:error dist} presents the distribution of the $1\sigma$ phase errors ($\Delta\phi$) extracted directly from the MCMC posterior distributions at each stage of the mock catalogue pipeline.

As expected, the idealised test-particle data exhibit reasonably small errors and are correlated with signal-to-noise ratio. The introduction of the `Finger of God' distance uncertainties selectively degrades the signal, driving up the phase errors particularly in regions where the intrinsic asymmetry amplitude is already weak. Finally, applying the non-uniform Gaia DR3 spatial footprint exacerbates these uncertainties. Mapping this degradation is critical, as it justifies the specific quality thresholds (e.g., masking fits with errors $>8.0^\circ$) implemented prior to the comparison of our simulated data with the Gaia data in Section \ref{final dev of gaia}.

\bsp	
\label{lastpage}
\end{document}